\documentclass[12pt]{article}
\usepackage{a4wide}
\usepackage{amsmath, color, cite}
\begin{document}
{\renewcommand{\thefootnote}{\fnsymbol{footnote}}
  \begin{center}
    {\LARGE  Deformed covariance in spherically symmetric\\
      vacuum models of loop quantum gravity:\\[2mm] Consistency in 
      Euclidean and self-dual gravity}\\ 
    \vspace{1.5em} Martin Bojowald$^1$\footnote{e-mail address: {\tt
        bojowald@gravity.psu.edu}}, Suddhasattwa Brahma$^2$\footnote{e-mail
      address: {\tt suddhasattwa.brahma@gmail.com}}, Ding
    Ding$^1$\footnote{e-mail address: {\tt dud79@psu.edu }}, Michele
    Ronco$^3$\footnote{e-mail address: {\tt michele.ronco@roma1.infn.it}}
    \\
     \vspace{0.5em}
    $^1$ Institute for Gravitation and the Cosmos,\\
    The Pennsylvania State
    University,\\
    104 Davey Lab, University Park, PA 16802, USA\\
    $^2$ Department of Physics, McGill University, \\
    Montr\'eal, QC H3A 2T8, Canada\\
    $^3$ Laboratoire de Physiqe Nucléaire et de Hautes Energies (LPNHE),\\
    Universit\'e Pierre et Marie Curie,\\
    Case courrier 200, 4 place Jussieu,   F-75005 Paris, France\
    \vspace{1.5em}
  
\end{center}
}

\setcounter{footnote}{0}

\begin{abstract}
  Different versions of consistent canonical realizations of hypersurface
  deformations of spherically symmetric space-times have been derived in
  models of loop quantum gravity, modifying the classical dynamics and
  sometimes also the structure of space-time. Based on a canonical version of
  effective field theory, this paper provides a unified treatment, showing
  that modified space-time structures are generic in this setting. The special
  case of Euclidean gravity demonstrates agreement also with existing operator
  calculations. 
\end{abstract}

\section{Introduction}

Several independent studies have shown that holonomy and inverse-triad
corrections from loop quantum gravity (LQG) modify hypersurface-deformation
brackets for spherically symmetric gravity and related midisuperspace models
\cite{JR,LTBII,ModCollapse,ModCollapse2,HigherSpatial,SphSymmOp,SphSymmCov,GowdyCov,MidiClass,BHSigChange},
thereby realizing a deformation of general covariance
\cite{Action,Normal,EffLine}. These modifications are closely related
\cite{DeformedCosmo} to anomaly-free models of perturbative cosmological
inhomogeneity constructed within the same framework
\cite{ConstraintAlgebra,ScalarGaugeInv,ScalarHol,ScalarTensorHol,ScalarHolInv},
suggesting that modified space-time structures may be a generic consequence of
quantum-geometry effects in loop quantum gravity.  In \cite{SphSymmComplex}
(see also \cite{SDGowdy}), however, it has been shown that such modifications
may be avoided if one uses self-dual connections and a densitized lapse
function, as in \cite{AshVar,SphKl1,SphKl2}, instead of real variables
\cite{AshVarReell}.  These models, valid for self-dual Lorentzian gravity with
Barbero--Immirzi parameter $\gamma=\pm i$ or Euclidean gravity with
Barbero--Immirzi parameter $\gamma=\pm 1$, are rather special because the
Hamiltonian constraint simplifies considerably compared with general
$\gamma$. It is therefore of interest to compare the structures encountered in
various models in order to determine whether undeformed space-time structures
could be realized more broadly.

Such a comparison is not obvious, for instance because the modifications
considered in \cite{SphSymmComplex} are different from those found in
anomaly-free models using real variables. In particular, those modifications
cannot be implemented in an anomaly-free manner for arbitrary choices of the
Barbero--Immirzi parameter: We will show that the classical form of the
constraint brackets can be retained only with a specific class of holonomy
modifications for $\gamma = \pm i$ (self-dual Lorentzian gravity) or
$\gamma=\pm 1$ (a special version of Euclidean gravity). More general
treatments of the self-dual or Euclidean case, implemented in close analogy
with the real connection formulation, lead to either anomalies or deformations
of the space-time structure.  This result then allows us to draw conclusions
about properties of the Hamiltonian constraint required for certain types of
modifications to be consistent.

At a technical level, an analysis of the Hamiltonian constraint and its
Poisson brackets indicates a formal relationship between modifications of
space-time structures and the appearance of spatial derivatives of the
densitized triads (canonically conjugate to the connection). Spatial
derivatives of the triad generically appear in the Hamiltonian constraints of
gravitational theories because they are required for curvature components. But
for $\gamma^2=\pm 1$, and \textit{only in this case}, they are completely absorbed in
the connection components through the spin connection which, in combination
with extrinsic-curvature components, forms the Ashtekar connection in the
self-dual case \cite{AshVar}, or the Ashtekar--Barbero connection in the real
case \cite{AshVarReell}.

This structural statement allows us to draw a first conclusion about the
genericness of modified space-time structures. Using standard arguments from
effective field theory (generalized here to a canonical setting), modified
brackets should be considered generic, unless one can show that the full
quantum theory has a symmetry that protects the derivative structure of terms
in the Hamiltonian constraint as encountered for self-dual variables, or more
generally for $\gamma^2=\pm 1$. No such symmetry is known. Although it has
been shown that the real Ashtekar--Barbero connection, unlike the self-dual
one, cannot be identified with the pull-back of a space-time connection, this
result is of an ``aesthetic nature'' \cite{GaugeConn} and does not
characterize the case of $\gamma^2=\pm 1$ via a physical symmetry that could
restrict possible quantum corrections. Moreover, applying this result in the
present context would amount to pre-supposing the classical space-time
structure in a model of quantum gravity. In canonical quantum gravity, the
structure of space-time is determined intrinsically, based on the observation
that space-time symmetries of a gravitational theory are gauge
transformations, generated in Hamiltonian form by the constraints that are to
be quantized in order to define canonical quantum gravity. Poisson brackets of
these constraints, or commutators of their operator versions, then encode the
structure of space-time. An analysis of possible consistent modifications of
these brackets, such that they remain closed but possibly with non-classical
structure functions, show whether the symmetries remain unviolated after
quantization. As we will see, such modifications with intact (but possibly
deformed) symmetry exist for any value of $\gamma$. Therefore, no value of
$\gamma$ is distinguished by the presence of a symmetry.

In this work, we will mainly focus on an interpretation of the constraints as
representing Euclidean gravity. We will then be exempt from having to consider
a possible role of reality conditions, the implementation of which remains
poorly understood in a quantum theory of self-dual variables. However, as the
constraints are formally identical in Euclidean gravity with $\gamma=\pm 1$
and self-dual Lorentzian gravity, our results can formally be used also in the
latter case.

\section{Unsolved Gauss constraint}

The model considered in \cite{SphSymmComplex}, following \cite{SphKl1},
consists of three canonical pairs of fields --- $A_i(x)$ and $E^i(x)$ for
$i=1,2,3$ depending on the radial coordinate $x$ of a spherically symmetric
manifold --- subject to three constraints. Two of the constraints function as
generators of hypersurface deformations in space-time and therefore encode the
structure of space-time. The third one, a Gauss constraint, implements an
internal symmetry of ${\rm SO}(2)$-rotations of two of the canonical pairs.

While the form of the Gauss constraint and the spatial generator of
hypersurface deformations (the diffeomorphism constraint) is strictly
determined by the canonical structure together with the corresponding Lie
algebras of infinitesimal rotations and 1-dimensional diffeomorphisms,
respectively, there is much freedom in specifying the normal generator of
hypersurface deformations, or the Hamiltonian constraint, even if the physical
dynamics is fixed. The version used in \cite{SphKl1,SphSymmComplex} is rather
special in that it is quadratic in the canonical fields and does not contain
spatial derivatives of $E^i$ (while first-order spatial derivatives of $A_i$
do appear). In the first part of this section we will strengthen the result of
\cite{SphSymmComplex} by showing that the consistent deformation found in this
paper is unique within a family of models that preserve the quadratic nature
and derivative structure of the Hamiltonian constraint. In the second part of
this section, however, we will show that this rigidity is not stable within a
larger class of models that determine the same classical dynamics but do not
respect the restricted derivative structure (parameterized by the so-called
Barbero--Immirzi paremeter $\gamma$ \cite{AshVarReell,Immirzi}). The following
sections will then place our discussion in a setting of effective field
theory, and highlight the role played by the Gauss constraint.

\subsection{Regaining the quadratic Hamiltonian constraint}  

In order to derive our rigidity result, we start from the condition that the
Poisson brackets of constraints be closed and see what kind of restrictions it
imposes on the form of constraints. The specific procedure follows the
classical (and classic) result \cite{Regained} that the full Hamiltonian
constraint, up to second order in derivatives, can be regained uniquely from
the classical hypersurface-deformation brackets, as specified in
\cite{DiracHamGR}. This procedure has already been applied to spherically
symmetric models in \cite{Action}, but only for modifications of the
dependence of the Hamiltonian constraint on the triad variables $E^i$.  Our
calculations here differ from \cite{Action} in that we use connection
variables $A_i$, and take into account potential modifications of the
dependence on these variables.

As already indicated, we assume for now that the Hamiltonian constraint is
quadratic in the canonical fields without spatial derivatives of the triad
$E^i$. This version of the constraint is realized in spherically symmetric
gravity if one uses self-dual connection variables \cite{AshVar} in Lorentzian
signature, or real Barbero-type variables \cite{AshVarReell} in Euclidean
signature such that the Barbero--Immirzi parameter is equal to $\gamma=\pm
1$. (One should also smear the Hamiltonian constraint with a lapse function of
density weight minus one to guarantee the quadratic nature.)  This parameter
is therefore fixed and does not appear in the remainder of this subsection.
Working with
\begin{equation} \label{PB1}
 \{A_1(x),E^1(y)\}= 2 G \delta(x,y)
\end{equation}
and
\begin{equation}\label{PB2}
\{A_2(x),E^2(y)\}= G \delta(x,y)\quad,\quad
\{A_3(x),E^3(y)\}= G \delta(x,y)
\end{equation}
while all other brackets of basic variables vanish.  (Note the missing factor
of $2$ in the last two brackets, compared with (\ref{PB1}), which is a
consequence of the fact that $(A_2,E^2)$ and $(A_3,E^3)$ encode the same
degree of freedom after the Gauss constraint is implemented.)
\begin{equation}
	\{A_1(x),E^1(y)\}=2\{A_{2/3}(x),E^{2/3}(y)\}=2\delta(x,y)\,.
\end{equation}
This canonical structure completely determines the Gauss constraint 
\begin{equation}
	G[\Lambda] = \frac{1}{2G}\int {\rm d}x
        \Lambda\left((E^1)'-2E^2A_3+2E^3A_2\right) 
\end{equation}
and the diffeomorphism constraint
\begin{equation}
	D[M] = \frac{1}{2G}\int{\rm d}x M\left(2A_3'E^3+2A_2'E^2-A_1(E^1)'\right)
\end{equation}
but not the Hamiltonian constraint.  Sometimes, it is convenient to combine
the diffeomorphism constraint $D[M]$ and the Gauss constraint $G[\Lambda]$ to
form the vector constraint
\begin{equation} \label{Vector}
	V[M] = D[M]+G[A_1M]= \frac{1}{G}\int{\rm d}x M\left((A_3'+A_1A_2)E^3+
   (A_2'-A_1A_3)E^2\right)\,.
\end{equation}

We will now use these constraints and attempt to derive the most general
form of the Hamiltonian constraint, purely quadratic in the canonical fields
and with up to first derivatives of $A_i$ but no derivatives of $E^i$, such
that all constraints have closed Poisson brackets.  With this assumption, we
can write the local (unsmeared) constraint as
\begin{equation} \label{HQuad}
 {\cal H}=H^{110}E^1E^2+H^{101}E^1E^3+H^{011}E^2E^3+H^{200}
 (E^1)^2+H^{020}(E^2)^2+H^{002}(E^3)^2\,,
\end{equation}
where we use the convention that $H[N]=(2G)^{-1} \int {\rm d}x N(x) {\cal H}$,
$H^{ijk}$ may be functions of $A_1$, $A_2$, $A_3$ and their spatial
derivatives up to first order.

\subsubsection{Diffeomorphism constraint}

We first consider the bracket of the Hamiltonian and diffeomorphism
constraints, writing it in local form as
\begin{eqnarray}
 \{{\cal H}(x),{\cal D}(y)\}&=& G\int{\rm d}z \left(2\frac{\delta
     {\cal H}(x)}{\delta A_1(z)}\frac{\delta {\cal D}(y)}{\delta
     E^1(z)}- 2\frac{\delta
     {\cal H}(x)}{\delta E^1(z)}\frac{\delta {\cal D}(y)}{\delta
     A_1(z)}\right.\\
 &&\qquad+ \left.\frac{\delta
     {\cal H}(x)}{\delta A_2(z)}\frac{\delta {\cal D}(y)}{\delta
     E^2(z)}- \frac{\delta
     {\cal H}(x)}{\delta E^2(z)}\frac{\delta {\cal D}(y)}{\delta
     A_2(z)}\right.\nonumber\\
 &&\qquad+ \left.\frac{\delta
     {\cal H}(x)}{\delta A_3(z)}\frac{\delta {\cal D}(y)}{\delta
     E^3(z)}- \frac{\delta
     {\cal H}(x)}{\delta E^3(z)}\frac{\delta {\cal D}(y)}{\delta
     A_3(z)}\right)\nonumber
\end{eqnarray}
where $D[M]=(2G)^{-1}\int{\rm d}x M(x){\cal D}(x)$.  If this bracket is to
correspond to classical hypersurface deformations, it should be equal to
\begin{equation} \label{HD}
 \{{\cal H}(x),{\cal D}(y)\}=2G\left( {\cal H}'(x)\delta(x,y)+2{\cal
   H}(x)\delta'(x,y))\right.\,,
\end{equation}
using the convention that a prime on a delta function always indicates a
derivative with respect to the first argument. Therefore,
\begin{equation} \label{deltaprime}
 \delta'(x,y)=-\delta'(y,x)\,.
\end{equation}
If the bracket is of the given form, the smeared constraints have the bracket
\begin{eqnarray}
	\{H[\underset{\sim}{N}],D[M]\}&=& \frac{1}{4G^2}\int{\rm d}x{\rm d}y
 \underset{\sim}{N}(x)M(y)\{{\cal    H}(x),{\cal 
   D}(y)\}\nonumber\\
&=&
\frac{1}{2G}\int{\rm d}x{\rm d}y \underset{\sim}{N}(x)M(y) \left((\partial_x{\cal
     H}(x))\delta(x,y)- 
   2{\cal H}(x) \partial_y\delta(x,y)\right)\nonumber\\
&=& -H[(\underset{\sim}{N}M)']+2H[\underset{\sim}{N}M']
=-H[M\underset{\sim}{N}'-M'\underset{\sim}{N}]
\end{eqnarray}
as required if $\underset{\sim}{N}$ has density weight minus one for
the purpose of having a quadratic Hamiltonian constraint.

We proceed by evaluating the Poisson bracket. Considering the assumed
dependence (\ref{HQuad}) of ${\cal H}$ on the canonical variables, we have
\begin{eqnarray}
  \{{\cal H}(x),{\cal D}(y)\} &=& 2G\int{\rm d}z \left(\left(\frac{\partial {\cal
          H}(x)}{\partial A_1(z)} \delta(x,z)+ \frac{\partial {\cal
          H}(x)}{A_1'(z)}\delta'(x,z)\right)
    \left(-A_1(y)\delta'(y,z)\right)\right.\nonumber\\
&&-\frac{\partial {\cal H}(x)}{\partial
      E^1(z)}\delta(x,z)\left(-(E^1)'(y)\delta(y,z)\right)\nonumber\\
&&    +\left(\frac{\partial {\cal
          H}(x)}{\partial A_2(z)} \delta(x,z)+ \frac{\partial {\cal
          H}(x)}{A_2'(z)}\delta'(x,z)\right) A_2'(y)\delta(y,z)\nonumber\\
&&-    \frac{\partial {\cal H}(x)}{\partial
      E^2(z)}\delta(x,z)E^2(y)\delta'(y,z)\nonumber\\
&&    +\left(\frac{\partial {\cal
          H}(x)}{\partial A_3(z)} \delta(x,z)+ \frac{\partial {\cal
          H}(x)}{A_3'(z)}\delta'(x,z)\right) A_3'(y)\delta(y,z)\nonumber\\
&&-\left.   \frac{\partial {\cal H}(x)}{\partial
      E^3(z)}\delta(x,z)E^3(y)\delta'(y,z)\right)\nonumber\\
  &=& 2G\left(\frac{\partial{\cal H}(x)}{\partial A_2(x)}A_2'(x)+
    \frac{\partial{\cal H}(x)}{\partial A_3(x)}A_3'(x)+ \frac{\partial {\cal
        H}(x)}{\partial E^1(x)}(E^1)'(x)\right)\delta(x,y)\nonumber\\
&&-
  \left(\frac{\partial{\cal H}(x)}{\partial A_1(x)}A_1(y)+ \frac{\partial{\cal
        H}(x)}{\partial E^2(x)}E^2(y)+ \frac{\partial {\cal H}(x)}{\partial
      E^3(x)}E^3(y)\right.\nonumber\\
&&\left. +\frac{\partial{\cal
        H}(x)}{\partial A_2'(x)} A_2'(y)+ \frac{\partial{\cal H}(x)}{\partial
      A_3'(x)}A_3'(y)\right) \delta'(y,x)\nonumber\\
&& -\int{\rm d}z \frac{\partial{\cal
      H}(x)}{\partial A_1'(z)}A_1(y) \delta'(x,z)\delta'(y,z)\,,
\end{eqnarray}
where we used (\ref{deltaprime}).

The last term has a product of two derivatives of delta functions, which does
not occur in (\ref{HD}). Integrating by parts can remove one of the
derivatives, but it also gives a second-order derivative of a delta function
which does not appear either in (\ref{HD}). The term, therefore, must be zero,
so that we already know that ${\cal H}$ cannot depend on $A_1'$. In order to
bring the remaining terms to a form close to (\ref{HD}), we use the identity
\begin{eqnarray}  \label{ABdeltaprime}
A(x)B(y)\delta'(y,x)&=&A(x)\partial_y\left(B(y)\delta(y,x)\right)-
A(x)B'(y)\delta(x,y)\nonumber\\
&=& A(x)\partial_y\left(B(x)\delta(y,x)\right)-
A(x)B'(x)\delta(x,y)\nonumber\\
&=& A(x)B(x)\delta'(y,x)-A(x)B'(x)\delta(x,y)
\end{eqnarray}
and write
\begin{eqnarray}
 \{{\cal H}(x),{\cal D}(y)\} &=& 2G \left(\frac{\partial{\cal H}(x)}{\partial
     A_1(x)}A_1'(x)+\frac{\partial{\cal H}(x)}{\partial A_2(x)}A_2'(x)+
    \frac{\partial{\cal H}(x)}{\partial A_3(x)}A_3'(x)\right.\nonumber\\
&&+ \frac{\partial{\cal
        H}(x)}{\partial A_2'(x)}A_2''(x)+\frac{\partial{\cal H}(x)}{\partial
      A_3'(x)}A_3''(x)\nonumber\\
&&\left.+ \frac{\partial {\cal
        H}(x)}{\partial E^1(x)}(E^1)'(x)+ \frac{\partial {\cal
        H}(x)}{\partial E^2(x)}(E^2)'(x)+ \frac{\partial {\cal
        H}(x)}{\partial E^3(x)}(E^3)'(x)\right)\delta(x,y)\nonumber\\
&&+ 2G\left(\frac{\partial{\cal H}(x)}{\partial
     A_1(x)}A_1(x)+\frac{\partial{\cal H}(x)}{\partial A_2'(x)}A_2'(x)+
    \frac{\partial{\cal H}(x)}{\partial A_3'(x)}A_3'(x)\right.\nonumber\\
&&\left. + \frac{\partial {\cal
        H}(x)}{\partial E^2(x)}E^2(x)+ \frac{\partial {\cal
        H}(x)}{\partial E^3(x)}E^3(x)\right)\delta'(x,y)\,.
\end{eqnarray}
Since ${\cal H}$ does not depend on $A_1'$, the first parenthesis (multiplied
by a delta function) is equal to ${\cal H}'$ without any further restriction
on the dependence on other canonical variables. In order to evaluate the
second parenthesis, which according to (\ref{HD}) should equal $4G{\cal H}$, we
use the quadratic form (\ref{HQuad}) and obtain the condition
\begin{eqnarray}
&&\frac{\partial{\cal H}(x)}{\partial
     A_1(x)}A_1(x)+\frac{\partial{\cal H}(x)}{\partial A_2'(x)}A_2'(x)+
    \frac{\partial{\cal H}(x)}{\partial A_3'(x)}A_3'(x)\nonumber\\
&&+
    H^{110}E^1E^2+H^{101}E^1E^3+
    2H^{011}E^2E^3+2H^{020}(E^2)^2+2H^{002}(E^3)^2 \nonumber\\
&=& 2\left(H^{110}E^1E^2+H^{101}E^1E^3+
    H^{011}E^2E^3+H^{020}(E^2)^2+H^{002}(E^3)^2\right)
\end{eqnarray}
or
\begin{eqnarray*}
\frac{\partial{\cal H}(x)}{\partial
     A_1(x)}A_1(x)+\frac{\partial{\cal H}(x)}{\partial A_2'(x)}A_2'(x)+
    \frac{\partial{\cal H}(x)}{\partial A_3'(x)}A_3'(x) 
= H^{110}E^1E^2+H^{101}E^1E^3+ 2H^{200}(E^1)^2
\end{eqnarray*}
after some cancellations. Comparing coefficients of $E^iE^j$ in this equation,
we obtain
\begin{eqnarray}
 \frac{\partial H^{110}}{\partial
     A_1}A_1+\frac{\partial H^{110}}{\partial A_2'}A_2'+
    \frac{\partial H^{110}}{\partial A_3'}A_3' &=& H^{110}\\
\frac{\partial H^{101}}{\partial
     A_1}A_1+\frac{\partial H^{101}}{\partial A_2'}A_2'+
    \frac{\partial H^{101}}{\partial A_3'}A_3' &=& H^{101}\\
\frac{\partial H^{011}}{\partial
     A_1}A_1+\frac{\partial H^{011}}{\partial A_2'}A_2'+
    \frac{\partial H^{011}}{\partial A_3'}A_3' &=& 0\\
 \frac{\partial H^{200}}{\partial
     A_1}A_1+\frac{\partial H^{200}}{\partial A_2'}A_2'+
    \frac{\partial H^{200}}{\partial A_3'}A_3' &=& 2H^{200}\\
 \frac{\partial H^{020}}{\partial
     A_1}A_1+\frac{\partial H^{020}}{\partial A_2'}A_2'+
    \frac{\partial H^{020}}{\partial A_3'}A_3' &=& 0\\
 \frac{\partial H^{002}}{\partial
     A_1}A_1+\frac{\partial H^{002}}{\partial A_2'}A_2'+
    \frac{\partial H^{002}}{\partial A_3'}A_3' &=& 0\,.
\end{eqnarray}
If we assume polynomial dependence of ${\cal H}$ on the connection variables,
we can conclude that the coefficients $H^{110}$ and $H^{101}$ must be linear
in $A_1$, $A_2'$ and $A_3'$, while $H^{200}$ must be quadratic in these
variables. The coefficients $H^{011}$, $H^{020}$ and $H^{002}$ cannot depend
on $A_1$, $A_2'$ or $A_3'$.

\subsubsection{Bracket of Hamiltonian constraints}

The Poisson bracket of two Hamiltonian constraints can be computed in a
similar way. Classically, we expect 
\begin{equation}
 \{{\cal H}(x),{\cal H}(y)\}= 2G\left(E^1(x)^2{\cal V}(x)\delta'(y,x)-
   E^1(y)^2{\cal 
   V}(y)\delta'(x,y)\right)
\end{equation}
with the local vector constraint ${\cal V}(x)$ such that
$V[M]=(2G)^{-1}\int{\rm d}xM(x){\cal V}(x)$. If the space-time structure is
deformed, the bracket is multiplied by a non-constant function $\beta$ which,
for a comparison with \cite{SphSymmComplex}, we assume to depend only on the
$A_i$. (This function should approach $\beta=1$ in some classical limit,
usually for small $A_i$.)  After using (\ref{HQuad}) and comparing
coefficients of $E^iE^j$, we obtain the equations
\begin{eqnarray}
 2\left(-2\frac{\partial H^{110}}{\partial A_1'}H^{200}- \frac{\partial
   H^{200}}{\partial A_1'}H^{110}\right)- \frac{\partial H^{110}}{\partial
   A_2'}H^{110} -2\frac{\partial H^{200}}{\partial A_2'}H^{020}-
 \frac{\partial H^{110}}{\partial A_3'}H^{101}- \frac{\partial
   H^{200}}{\partial A_3'}H^{011}&& \label{1}\\
 = 4 \beta (A_2'-A_1A_3)&&\nonumber\\
 2\left(-2\frac{\partial H^{101}}{\partial A_1'}H^{200}- \frac{\partial
   H^{200}}{\partial A_1'}H^{101}\right)- \frac{\partial H^{101}}{\partial
   A_2'}H^{110} -2\frac{\partial H^{200}}{\partial A_2'}H^{011}-
 \frac{\partial H^{101}}{\partial A_3'}H^{101}- \frac{\partial
   H^{200}}{\partial A_3'}H^{002}&&\label{2}\\
 = 4\beta (A_3'+A_1A_2)\,,
\end{eqnarray}
which are sensitive to the modification function $\beta$, as
well as several $\beta$-independent equations:
\begin{eqnarray}
4\frac{\partial H^{200}}{\partial A_1'}H^{200}+ \frac{\partial
  H^{200}}{\partial A_2'} H^{110}+ \frac{\partial H^{200}}{\partial
  A_3'}H^{101} =0&&\label{3}\\
2\left(\frac{\partial H^{110}}{\partial A_1'}H^{110}+2 \frac{\partial
  H^{020}}{\partial A_1'} H^{200}\right)+ 2\frac{\partial H^{110}}{\partial
  A_2'}H^{020} + \frac{\partial H^{020}}{\partial A_2'}H^{110}+ \frac{\partial
  H^{110}}{\partial A_3'} H^{011}+ \frac{\partial H^{020}}{\partial
  A_3'}H^{101} =0&&\label{8}\\
2\left(\frac{\partial H^{101}}{\partial A_1'}H^{101}+ 2\frac{\partial
  H^{002}}{\partial A_1'} H^{200}\right)+ \frac{\partial H^{101}}{\partial
  A_2'}H^{011} + \frac{\partial H^{002}}{\partial A_2'}H^{110}+ 2\frac{\partial
  H^{101}}{\partial A_3'} H^{002}+ \frac{\partial H^{002}}{\partial
  A_3'}H^{101} =0&&\label{9}\\
2\left(2\frac{\partial H^{011}}{\partial A_1'}H^{200}+ \frac{\partial
  H^{101}}{\partial A_1'} H^{110}+ \frac{\partial H^{110}}{\partial
  A_1'}H^{101}\right)&&\nonumber\\
 + \frac{\partial H^{011}}{\partial
  A_2'}H^{110} + 2\frac{\partial H^{101}}{\partial A_2'}H^{020}+\frac{\partial
  H^{110}}{\partial A_2'}H^{011}+\frac{\partial
  H^{011}}{\partial A_3'} H^{101}+ \frac{\partial H^{101}}{\partial
  A_3'}H^{011}+2\frac{\partial H^{110}}{\partial A_3'}H^{002} =0 &&\,.\label{10}
\end{eqnarray}
Four additional equations,
\begin{eqnarray}
2\frac{\partial H^{020}}{\partial A_1'}H^{110}+ 2\frac{\partial
  H^{020}}{\partial A_2'} H^{020}+ \frac{\partial H^{020}}{\partial
  A_3'}H^{011} =0&&\label{4}\\
2\frac{\partial H^{002}}{\partial A_1'}H^{101}+ \frac{\partial
  H^{002}}{\partial A_2'} H^{011}+ 2\frac{\partial H^{002}}{\partial
  A_3'}H^{002} =0&&\label{5}\\
2\left(\frac{\partial H^{011}}{\partial A_1'}H^{110}+ \frac{\partial
  H^{020}}{\partial A_1'} H^{101}\right)+ 2\frac{\partial H^{011}}{\partial
  A_2'}H^{020} + \frac{\partial H^{020}}{\partial A_2'}H^{011}+ \frac{\partial
  H^{011}}{\partial A_3'} H^{011}+ 2\frac{\partial H^{020}}{\partial
  A_3'}H^{002} =0&&\label{6}\\
2\left(\frac{\partial H^{011}}{\partial A_1'}H^{101}+ \frac{\partial
  H^{002}}{\partial A_1'} H^{110}\right)+ \frac{\partial H^{011}}{\partial
  A_2'}H^{011} + 2\frac{\partial H^{002}}{\partial A_2'}H^{020}+ 2\frac{\partial
  H^{011}}{\partial A_3'} H^{002}+ \frac{\partial H^{002}}{\partial
  A_3'}H^{011} =0&&\label{7}
\end{eqnarray}
are identically satisfied, given that $H^{011}$, $H^{020}$ and $H^{002}$
cannot depend on $A_i'$. 
Because ${\cal H}$ cannot depend on $A_1'$, we may simplify the set of
equations to
\begin{eqnarray}
 - \frac{\partial H^{110}}{\partial
   A_2'}H^{110} -2\frac{\partial H^{200}}{\partial A_2'}H^{020}-
 \frac{\partial H^{110}}{\partial A_3'}H^{101}- \frac{\partial
   H^{200}}{\partial A_3'}H^{011}&=&4 \beta (A_2'-A_1A_3) \label{1b}\\
- \frac{\partial H^{101}}{\partial
   A_2'}H^{110} -2\frac{\partial H^{200}}{\partial A_2'}H^{011}-
 \frac{\partial H^{101}}{\partial A_3'}H^{101}- \frac{\partial
   H^{200}}{\partial A_3'}H^{002}&=&4\beta (A_3'+A_1A_2)\label{2b}\\
\frac{\partial
  H^{200}}{\partial A_2'} H^{110}+ \frac{\partial H^{200}}{\partial
  A_3'}H^{101} &=&0\label{3b}\\
2\frac{\partial H^{110}}{\partial
  A_2'}H^{020} + \frac{\partial
  H^{110}}{\partial A_3'} H^{011} &=&0\label{8b}\\
\frac{\partial H^{101}}{\partial
  A_2'}H^{011} + 2\frac{\partial
  H^{101}}{\partial A_3'} H^{002}&=&0\label{9b}\\
2\frac{\partial H^{101}}{\partial A_2'}H^{020}+\frac{\partial
  H^{110}}{\partial A_2'}H^{011}+ \frac{\partial H^{101}}{\partial
  A_3'}H^{011}+2\frac{\partial H^{110}}{\partial A_3'}H^{002} &=&0 \,.\label{10b}
\end{eqnarray}

\subsubsection{Gauss constraint}

The Gauss constraint further restricts the combinations of basic variables
which can appear in the Hamiltonian constraint. The gauge-invariant
combinations that contribute to the classical constraint are $E^1$,
$(E^2)^2+(E^3)^2$, $A_2E^2+A_3E^3$, $A_2^2+A_3^2$ and
$A_1(A_2E^2+A_3E^3)-(A_2'E^3-A_2'E^2)$. (The identity (\ref{ABdeltaprime}) is
useful for seeing that the last combination has a vanishing Poisson bracket
with the unsmeared Gauss constraint.)  These expressions show that $A_1$,
$A_2'$ and $A_3'$ can appear in gauge-invariant form only in combination with
$E^2$ and $E^3$. It is therefore impossible to fulfill the condition that
$H^{200}$ be quadratic in $A_1$, $A_2'$ and $A_3'$ because $H^{200}$ is
defined as the $E$-independent coefficient of $(E^1)^2$ in the Hamiltonian
constraint. For Hamiltonian constraints quadratic in $E^i$, we have
$H^{200}=0$. 

Equations (\ref{1b}) and (\ref{2b}) then simplify to
\begin{eqnarray}
 - \frac{\partial H^{110}}{\partial
   A_2'}H^{110} -
 \frac{\partial H^{110}}{\partial A_3'}H^{101} &=& 4\beta (A_2'-A_1A_3) \\
- \frac{\partial H^{101}}{\partial
   A_2'}H^{110} -
 \frac{\partial H^{101}}{\partial A_3'}H^{101}&=&  4\beta (A_3'+A_1A_2)\,.
\end{eqnarray}
For $\beta=1$, these equations are obeyed by the classical $H^{110}_{\rm cl}=
2(A_1A_2+A_3')$ and $H^{101}_{\rm cl}=2(A_1A_3-A_2')$, as they should. For
$\beta\not=1$, we can solve these two equations by
$H^{110}=\beta_1H^{110}_{\rm cl}$ and $H^{101}=\beta_2H^{101}_{\rm cl}$,
provided that $\beta_1$ and $\beta_2$ do not depend on spatial derivatives of
$A_i$ and are such that $\beta_1\beta_2=\beta$. Invariance under
transformations generated by the Gauss constraint, which mix the terms of
$H^{110}_{\rm cl}$ and $H^{101}_{\rm cl}$, implies that $\beta_1=\beta_2$, and
therefore $\beta>0$ and $\beta_1=\beta_2=\sqrt{\beta}$.  This modification
function can be eliminated from the contributions of $H^{110}$ and $H^{101}$
to the constraint by absorbing it in the lapse function, thus moving the
modification to the remaining contributions from
$H^{020}=\beta^{-1/2}H^{020}_{\rm cl}$ and $H^{002}=\beta^{-1/2}H^{002}_{\rm
  cl}$. Therefore, the only non-trivial modification of the dynamics is in
the contributions from $H^{020}$ and $H^{002}$ which, as already shown, can
only depend on $A_2$ and $A_3$. Again invoking transformations generated by
the Gauss constraint, the modified term $\beta^{-1/2}(H^{020}_{\rm
  cl}+H^{002}_{\rm cl})$ is an arbitrary (positive) function of $A_2^2+A_3^2$,
which is equivalent to the modification found in \cite{SphSymmComplex} and
therefore strengthens their result.

If we relax the condition that the Hamiltonian constraint not depend on
spatial derivatives of the densitized triad, additional gauge invariant
combinations are possible. For instance, the extrinsic-curvature component
\begin{equation}
 K_1=A_1-\frac{(E^2)'E^3-E^2(E^3)'}{(E^2)^2+(E^3)^2}
\end{equation}
is gauge invariant. Moreover, if spatial derivatives of the densitized triad
are allowed, the Gauss constraint can be used to rewrite the Hamiltonian
constraint without changing the on-shell behavior. For instance, the identity
\begin{equation}
 A_1(A_2E^2+A_3E^3)+2E^2A_3'-2E^3A_2' =  (E^1)''+ A_2(A_1E^2+2(E^3)')+
 A_3(A_1E^3-2(E^2)')-{\cal G}'
\end{equation}
eliminates spatial derivatives of $A_2$ and $A_3$ from the
Hamiltonian constraint, in favor of a second-order spatial derivative of
$E^1$. This new form is much closer to the expression of the Hamiltonian
constraint in extrinsic-curvature variables \cite{SphSymmHam}, and may allow
different modified brackets than the quadratic version (\ref{HQuad}) even if
one works with the reduced Ashtekar connections $A_i$.

The possibility of rewriting the Hamiltonian constraint by using the Gauss
constraint explains why different formulations of the same classical theory
may give rise to different modified brackets: The Gauss constraint depends on
$A_2$ and $A_3$, and therefore, depending on how it is used in writing the
Hamiltonian constraint, it restricts possible modifications.  In
extrinsic-curvature variables, this ambiguity does not appear because the
Gauss constraint is solved explicitly.

From the perspective of effective field theory, applied here to
  the classical structure of up to second-order derivatives, restricting the
dependence of the Hamiltonian constraint on spatial derivatives of $E^i$ leads
to non-generic models. The classical constraint is quadratic in $A_i$, which,
according to the field equations implied by the theory, amounts to terms with
up to two derivatives. Any term that is consistent with the symmetries of the
theory (generated by the constraints) and has up to two derivatives (temporal
or spatial) should then be allowed for a generic model. Such theories should
include terms with up to second-order spatial derivatives of $E^i$, in
addition to the quadratic terms in $A_i$ which contribute two time
derivatives. (A higher-derivative theory beyond second order
  would be obtained by including quantum back-reaction effects, which is not
  the purpose of this paper.)

\subsection{Arbitrary Barbero--Immirzi parameter}

We will now show that the preceding rigidity result is not stable within a
class of models in which spatial derivatives of the densitized triad are
allowed to appear. A suitable set of constraints that describes the same
classical physics as, depending on the signature, Euclidean or self-dual
gravity is obtained by letting the Barbero--Immirzi parameter vary, instead of
fixing it to a specific value such that $\gamma^2=\pm1$. The modification
found in \cite{SphSymmComplex} is therefore not generic. To this end, we will
now switch to a general setting of spherically symmetric gravity in which the
Barbero--Immirzi parameter and other numerical factors (as well as the
gravitational constant $G$) are included.

Spherically symmetric gravity can be formulated as a Hamiltonian theory with
phase space given by the canonical pairs, subject to three constraints. This
setting has been formulated in \cite{SphKl1} for self-dual variables and in
\cite{SphSymmHam} for real variables. In order to avoid having to impose
reality conditions, we follow the latter notation, in which the canonical
pairs $(A_1,E^1)$, $(A_2,E^2)$ and
$(A_3,E^3)$ are such that
\begin{equation}
 \{A_1(x),E^1(y)\}= 2\gamma G \delta(x,y)
\end{equation}
and
\begin{equation}
\{A_2(x),E^2(y)\}= \gamma G \delta(x,y)\quad,\quad
\{A_3(x),E^3(y)\}= \gamma G \delta(x,y)
\end{equation}
(a version of (\ref{PB1}) and (\ref{PB2}) for arbitrary real $\gamma$).  They
are subject to the Gauss constraint
\begin{equation} \label{Gauss} 
G[\Lambda] = \frac{1}{2\gamma G} \int {\rm d}x
  \Lambda \left((E^1)' + 2A_2 E^3 - 2A_3 E^2\right)
\end{equation}
smeared with a multiplier $\Lambda$, the diffeomorphism constraint
\begin{equation}
  D[N^x] = \frac{1}{2\gamma G} 
\int {\rm d}x\,  N^x \, \left(-A_1 (E^1)' + 2A'_3 E^3 + 2A'_2
    E^2 \right) 
\end{equation}
smeared with the shift vector $N^x$, and the Hamiltonian constraint
\begin{eqnarray}
\label{cartham}
H[\underset{\sim}{N}] &=& \frac{1}{2 G}
 \int {\rm d}x\, \underset{\sim}{N}\, \left(2 A_1 E^1
  (A_2 E^2 + A_3 
  E^3)\right.\nonumber\\
&& + (A_2^2 + A_3^2-1)
\left((E^2)^2+(E^3)^2\right)+ 2 E^1 \left(E^2 A'_3 - E^3 A'_2\right)\nonumber\\
&&+ \left.(\epsilon-\gamma^2)\left(2K_1 E^1 (K_2 E^2 + K_3 E^3) + ((K_2)^2 +
(K_3)^2)((E^2)^2+(E^3)^2)\right)\right)\\
& =& H^{\rm E}[\underset{\sim}{N}] + H^{\rm L}[\underset{\sim}{N}]\nonumber
\end{eqnarray}
smeared with the lapse function $\underset{\sim}{N}$ of density weight
$-1$. The non-polynomial relationship between the extrinsic-curvature
components $K_1$, $K_2$ and $K_3$ with the basic variables is given below. 

In all three constraints, the prime represents a derivative with respect to
the radial coordinate $x$.  Moreover, $\gamma$ in (\ref{cartham}) is the
Barbero--Immirzi parameter \cite{AshVarReell,Immirzi} and $\epsilon=\pm 1$ the
space-time signature, such that $\epsilon=1$ in the Euclidean case and
$\epsilon=-1$ in the Lorentzian case.  As usual, it is convenient to split the
Hamiltonian constraint into the Euclidean part
\begin{eqnarray}
 H^{\rm E}[ \underset{\sim}{N}] &=& \frac{1}{2 G}
 \int {\rm d}x\, \underset{\sim}{N}\, \left(2 A_1 E^1
  (A_2 E^2 + A_3 
  E^3)\right.\nonumber\\
&& + \left. (A_2^2 + A_3^2-1)
\left((E^2)^2+(E^3)^2\right)+ 2 E^1 \left(E^2 A'_3 - E^3 A'_2\right)\right)
\end{eqnarray}
and  the ``Lorentzian'' contribution
\begin{eqnarray}
 H^{\rm L}[ \underset{\sim}{N}] &=& -\frac{\gamma^2-\epsilon}{2 G}
 \int {\rm d}x\, \underset{\sim}{N}\, 
   \left(2K_1 E^1 (K_2 E^2 + K_3 E^3) + ((K_2)^2 + 
(K_3)^2)((E^2)^2+(E^3)^2)\right)\,.
\end{eqnarray}
Thus, $H[\underset{\sim}{N}]=H^{\rm E}[ \underset{\sim}{N}]$ for $\gamma=\pm 1$
in Euclidean signature ($\epsilon=1$), while the ``Lorentzian'' contribution
(a slight misnomer) also contributes in Euclidean signature if $\gamma\not=\pm
1$. (The Lorentzian contribution is always required in Lorentzian signature if
one works with real $\gamma$ such that the Poisson brackets are real.)
The canonical variables $A_1$, $E^2$ and $E^3$ have density weight one.

The geometrical meaning of the phase-space variables is determined as follows:
The fields $E^1$, $E^2$ and $E^3$, as the components of a spherically
symmetric densitized triad, describe a spatial metric $q_{ab}$ according to
the line element
\begin{equation} \label{ds}
 {\rm d}s^2=q_{ab}{\rm d}x^a{\rm d}x^b=
 \frac{(E^2)^2+(E^3)^2}{|E^1|} {\rm d}x^2+ |E^1| ({\rm
   d}\vartheta^2+ \sin^2\vartheta{\rm d}\varphi^2)\,.
\end{equation}
The densitized triad also determines a spin connection such that it is
constant with respect to the resulting covariant derivative. The components of
this spin connection are functions of the densitized triad and its first
spatial derivatives:
\begin{equation}\label{spinconn}
  \Gamma_1 = \frac{E^3 (E^2)' - E^2 (E^3)'}{(E^2)^2+(E^3)^2}\quad, \quad
  \Gamma_2 = 
  -\frac{1}{2} \frac{(E^1)' E^3}{(E^2)^2+(E^3)^2}\quad, \quad \Gamma_3 =
  \frac{1}{2} \frac{(E^1)'  E^2}{(E^2)^2+(E^3)^2}\,.
\end{equation}
The densitized triad is canonically conjugate to components of extrinsic
curvature, $K_i$, $i=1,2,3$. Since the $\Gamma_i$ depend only on $E^i$, one
can add them to $K_i$ without changing the latter's canonical relationships
with $E^i$. In this way, the canonical connection components
$A_i=\Gamma_i+\gamma K_i$ are obtained, using the Barbero--Immirzi parameter
$\gamma$. 

The constrained system is first class, with brackets of the constraints
$D[N^x]$ and $H[\underset{\sim}{N}]$ according to Dirac's hypersurface
deformations \cite{DiracHamGR} (taking into account the density weight of
$\underset{\sim}{N}$ in the Hamiltonian constraint used here). In particular,
the bracket $\{ H[ \underset{\sim}{N}], H[ \underset{\sim}{M}]\}$ should be
proportional to the diffeomorphism constraint, up to possible contributions
from the Gauss constraint.  We display the relevant derivations in a more
general setting, following the observation \cite{SphSymmComplex} that, for
$\gamma^2=\epsilon$, the constraint brackets remain closed in the presence of
a ``magnetic-field'' modification, replacing $B_1:=A_2^2 + A_3^2-1$ in the
Euclidean part of the Hamiltonian constraint with an arbitrary function
$f(A_2^2 + A_3^2-1)$. Our aim is to determine whether this modification can be
carried over to the Lorentzian contribution.

We begin with the bracket of two modified Euclidean parts, $\{H^{\rm
  E}[\underset{\sim}{N}], H^{\rm E}[\underset{\sim}{M}]\}$. Thanks to
antisymmetry of the bracket in $\underset{\sim}{N}$ and $\underset{\sim}{M}$,
we need consider only those brackets of terms that lead to derivatives of
delta functions. There are two such contributions,
\begin{eqnarray}
 &&\{2A_1(x)E^2(x)(A_2(x)E^2(x)+A_3(x)E^3(x)),
 2E^1(y)(E^2(y)A_3(y)'-E^3(y)A_2(y)')\}\nonumber\\
&=& (\cdots)\delta(x,y)-
 4\gamma G A_1(x)E^1(x)E^1(y) \left(A_3(x)E^2(y)-
   A_2(x)E^3(y)\right) \partial_y\delta(x,y)
\end{eqnarray}
and
\begin{eqnarray}
 && \{2E^1(x)(E^2(x)A_3(x)'-E^3(x)A_2(x)'),
 2E^1(y)(E^2(y)A_3(y)'-E^3(y)A_2(y)')\}\nonumber\\
&=& (\cdots)\delta(x,y)-
4\gamma G E^1(x)E^1(y)\left(\left(E^2(x)A_2(y)'+E^3(x)A_3(y)'\right) 
\partial_x\delta(x,y)\right.\nonumber\\ 
&&-
\left.\left(E^2(y)A_2(x)'+E^3(y)A_3(x)'\right) \partial_y\delta(x,y)\right) \,.
\end{eqnarray}
With these two ingredients, we obtain
\begin{eqnarray}\label{euclhyp}
 \{H^{\rm
  E}[\underset{\sim}{N}], H^{\rm E}[\underset{\sim}{M}]\}&=&
\frac{\gamma}{ G} \int {\rm d}x
\left(\underset{\sim}{N}'\underset{\sim}{M}-
\underset{\sim}{N}\underset{\sim}{M}'\right)
  (E^1)^2 
  \left(A_1(A_2E^3-A_3E^2)+ E^2A_2'+E^3A_3'\right)\nonumber\\
&=& \gamma^2 V[(E^1)^2(\underset{\sim}{N}'\underset{\sim}{M} -
\underset{\sim}{M}'\underset{\sim}{N})] 
\end{eqnarray}
where
\begin{equation}
V[\Lambda] =\frac{1}{\gamma G} \int {\rm d}x\,  \Lambda \, \left(A_1 (E^2 A_3
  - E^3   A_2) + A'_3 E^3 + A'_2 
E^2 \right) 
\end{equation}
is the vector constraint constraint (\ref{Vector}), $V[\Lambda] = D[\Lambda] +
G[A_1\Lambda]$, related to the diffeomorphism constraint $D$ through a
contribution from the Gauss constraint (\ref{Gauss}).

Using $\sqrt{\det{q}}=\sqrt{|E^1|((E^2)^2+(E^3)^2)}$ from (\ref{ds}), we can
write the smearing function in (\ref{euclhyp}) as
\begin{equation} \label{smearing}
  (E^1)^2\left(\underset{\sim}{N}'\underset{\sim}{M} -
  \underset{\sim}{M}'\underset{\sim}{N}\right) = \frac{|E^1|}{(E^2)^2+(E^3)^2}
(N'M-M'N) 
\end{equation}
where $N=\sqrt{|E^1|((E^2)^2+(E^3)^2)}\underset{\sim}{N}$ and
$M=\sqrt{|E^1|((E^2)^2+(E^3)^2)} \underset{\sim}{M}$ are lapse functions
without density weight. The coefficient $|E^1|/\left((E^2)^2+(E^3)^2\right)$
in (\ref{smearing}) is, according to (\ref{ds}), the radial component of the
inverse spatial metric, in agreement with the classical form of
hypersurface-deformation brackets.  The system is therefore anomaly-free for
any modification $f$ in (\ref{cartham}) without any modification of the
constraint brackets and the space-time structure --- provided the Lorentzian
part does not contribute to the Hamiltonian constraint, that is in Euclidean
gravity with $\gamma=\pm 1$ or in Lorentzian gravity with $\gamma=\pm i$. This
is consistent with the results reported in \cite{SphSymmComplex}.

It is easy to see that any function $f(A_2^2+A_3^2-1)$ can be used in the
modified Euclidean part because this term does not produce derivatives of
delta functions in the Poisson bracket of two Euclidean constraints. Moreover,
because $A_2$ and $A_3$ are scalars without density weight, any such term has
the correct Poisson bracket with the diffeomorphism constraint. However, if
$\gamma^2 \neq \epsilon$, the cross-term $\{H^{\rm E}[\underset{\sim}{N}],
H^{\rm L}[\underset{\sim}{M}]\}$ in the Poisson bracket of two Hamiltonian
constraints does receive a contribution from $f(A_2^2+A_3^2-1)$ in $H^{\rm
  E}[\underset{\sim}{N}]$ because $H^{\rm L}[\underset{\sim}{M}]$, written in
the canonical variables $A_i$ and $E^i$, contains spatial derivatives of $E^i$
through $\Gamma_i$. An explicit calculation is therefore required to check
whether the bracket can still be closed for $f(A_2^2+A_3^2-1)\not=
A_2^2+A_3^2-1$. 

We first compute The Poisson brackets of each individual term in $H^{\rm
  E}[\underset{\sim}{N}]$ with the full $H^{\rm L}[\underset{\sim}{M}]$:
We obtain
\begin{eqnarray}
&& \frac{1}{ G}\{\smallint{\rm d}x N(x)
A_1(x)E^1(x)(A_2(x)E^2(x)+A_3(x)E^3(x)), H^{\rm
  L}[\underset{\sim}{M}]\}\nonumber\\  
 &=& \frac{\gamma^2-\epsilon}{2\gamma^2G^2} \int{\rm d}x{\rm d}yN(x)M(y)
 \left((\cdots)\delta(x,y)\right.\nonumber\\
&&\qquad-2 A_1(x)E^1(x)E^1(y) (A_2(y)E^2(y)+A_3(y)E^3(y))
   \{A_2(x)E^2(x)+A_3(x)E^3(x), \Gamma_1(y)\}\nonumber\\
&&\qquad+ E^1(x) (A_2(x)E^2(x)+A_3(x)E^3(x))
   \left(E^2(y)^2+E^3(y)^2\right)\nonumber\\
&&\qquad\qquad\times \left. \{A_1(x),
   -2(A_2(y)\Gamma_2(y)+A_3(y)\Gamma_3(y))+
   \Gamma_2(y)^2+\Gamma_3(y)^2\}\right)\nonumber\\
&=& \frac{\gamma^2-\epsilon}{2\gamma G} \int{\rm d}x{\rm d}yN(x)M(y) \Bigl(
  -2A_1(x)E^1(x)E^1(y) (A_2(y)E^2(y)\nonumber\\
&&\qquad+A_3(y)E^3(y))
  \frac{E^2(x)E^3(y)-E^2(y)E^3(x)}{E^2(y)^2+E^3(y)^2}\nonumber\\
&&\qquad +2 E^1(x)
  (E^2(y)^2+E^3(y)^2) (A_2(x)E^2(x)+A_3(x)E^3(x))\nonumber\\
&&\qquad\qquad\times  \frac{A_2(y)E^3(y)-A_3(y)E^2(y)- E^3(y)\Gamma_2(y)+
    E^2(y)\Gamma_3(y)}{E^2(y)^2+E^3(y)^2}\Bigr) \partial_y\delta(x,y)\nonumber\\
&=& -\frac{\gamma^2-\epsilon}{2\gamma G}\int{\rm d}xN(x) M'(x) E^1
(A_2E^2+A_3E^3) 
\left((E^1)'+2A_2E^3- 2A_3E^2\right)\nonumber\\
&=& - (\gamma^2-\epsilon) G[NM' E^1 (A_2E^2+A_3E^3)]
\end{eqnarray}
up to terms that cancel out when inserted in the antisymmetric $\{H^{\rm
  E}[\underset{\sim}{N}], H^{\rm L}[\underset{\sim}{M}]\}+ \{H^{\rm
  L}[\underset{\sim}{N}], H^{\rm E}[\underset{\sim}{M}]\}$. In the detailed
calculations, we have used the explicit expressions for the $\Gamma_i$, from
which we also obtain the useful identity
\begin{equation}
 \gamma(K_2E^2+K_3E^3)=A_2E^2+A_3E^3
\end{equation}
because $\Gamma_2E^2+\Gamma_3E^3$ is identically zero.

The second term,
\begin{eqnarray}
 && \frac{1}{2 G}\{\smallint{\rm d}x N(x)
f(A_2(x)^2+A_3(x)^2-1)(E^2(x)^2+E^3(x)^2), H^{\rm
  L}[\underset{\sim}{M}]\}\nonumber\\  
 &=& \frac{\gamma^2-\epsilon}{2\gamma^2G^2} \int{\rm d}x{\rm d}yN(x)M(y)
 \Bigl((\cdots)\delta(x,y)\nonumber\\
&&\qquad -2 \dot{f}(x) (E^2(x)^2+E^3(x)^2) E^1(y) (A_2(y)E^2(y)+A_3(y)E^3(y))
\{A_2(x)^2+A_3(x)^2,\Gamma_1(y)\}\Bigr)\nonumber\\
&=& \frac{\gamma^2-\epsilon}{2\gamma G} \int{\rm d}x{\rm d}yN(x)M(y)
 \Bigl((\cdots)\delta(x,y)\nonumber\\
&&\qquad -2 \dot{f}(x) (E^2(x)^2+E^3(x)^2) E^1(y)
(A_2(y)E^2(y)+A_3(y)E^3(y))\nonumber\\
&&\qquad\qquad\times
\frac{2A_2(x)E^3(y)-A_3(x)E^2(y)}{E^2(y)^2+E^3(y)^2}\partial_y\delta(x,y)
\Bigr)\nonumber\\ 
&=& 2(\gamma^2-\epsilon)
G[NM'\dot{f}E^1(A_2E^2+A_3E^3)]- \frac{\gamma^2-\epsilon}{2\gamma G} \int{\rm
  d}xNM' \dot{f} 
E^1(E^1)' (A_2E^2+A_3E^3)\,, \label{brack1}
\end{eqnarray}
does not vanish on the constraint surface. Therefore, the function $f$, whose
derivative by its argument we have denoted by $\dot{f}$, is now relevant for
closed brackets. In particular, the last contribution containing $(E^1)'$ must
be canceled by a corresponding term in the remaining bracket.

In this last bracket,
\begin{eqnarray}
 B&:=& \frac{1}{G}\{\smallint{\rm d}x N(x)
E^1(x)(E^2(x)A_3(x)'-E^3(x)A_2(x)'), H^{\rm
  L}[\underset{\sim}{M}]\}\nonumber\\  
 &=& \frac{\gamma^2-\epsilon}{2\gamma^2G^2} \int{\rm d}x{\rm d}yN(x)M(y)
 \Bigl((\cdots)\delta(x,y)\nonumber\\
&&\qquad +2 E^1(x)E^1(y) (A_2(y)E^2(y)+A_3(y)E^3(y))
\{E^2(x)A_3(x)'-E^3(x)A_2(x)',-\Gamma_1(y)\}\nonumber\\
&&\qquad+ 2E^1(x)E^1(y) (A_1(y)-\Gamma_1(y))\{E^2(x)A_3(x)'-E^3(x)A_2(x)',
A_2(y)E^2(y)+A_3(y)E^3(y)\}\nonumber\\
&&\qquad -2 E^1(x) (E^2(y)^2+E^3(y)^2) \bigl((A_2(y)-\Gamma_2(y))
  \{E^2(x)A_3(x)'-E^3(x)A_2(x)', \Gamma_2(y)\}\nonumber\\
&&\qquad\qquad +
  (A_3(y)-\Gamma_3(y))\{E^2(x)A_3(x)'-E^3(x)A_2(x)', \Gamma_3(y)\}\bigr)
\Bigr)\nonumber\\
&=& \frac{\gamma^2-\epsilon}{2\gamma G} \int{\rm d}x{\rm d}yN(x)M(y)
 \Bigl((\cdots)\delta(x,y)\nonumber\\
&&\qquad -2E^1(x)E^1(y) (A_2(y)E^2(y)+A_3(y)E^3(y))
\frac{E^2(x)E^2(y)'+E^3(x)E^3(y)'}{E^2(y)^2+E^3(y)^2} 
\partial_x\delta(x,y)\nonumber\\
&& \qquad +2
E^1(x)E^1(y) (A_2(y)E^2(y)+A_3(y)E^3(y))
\frac{E^2(x)E^3(y)+E^3(x)E^2(y)}{E^2(y)^2+E^3(y)^2} 
\partial_x\partial_y\delta(x,y)\nonumber\\ 
&&\qquad +2(A_1(y)-\Gamma_1(y)) E^1(x)E^1(y)
(E^2(x)A_3(y)-E^3(x)A_2(y)) \partial_x\delta(x,y)\nonumber\\
&&\qquad + E^1(x)E^1(y)
\left((A_2(y)-\Gamma_2(y))E^2(x)+
  (A_3(y)-\Gamma_3(y))E^3(y)\right) \partial_x\delta(x,y)\Bigr) \,,
\end{eqnarray}
we have a contribution from a second-order derivative of the delta
function. Integrating by parts once in this term and taking into account its
contributions to $NM'$ and $N'M$, respectively, (noting that terms with $N'M'$
cancel out in the final antisymmetric bracket) we write
\begin{eqnarray}
B &=& \frac{\gamma^2-\epsilon}{2\gamma G} \int{\rm d}x{\rm d}yN(x)M(y)
 \Biggl((\cdots)\delta(x,y)\nonumber\\
&&\qquad -2\frac{E^1(x)E^1(y)}{E^2(y)^2+E^3(y)^2}
\Biggl((E^2(x)E^2(y)'+E^3(x)E^3(y)') (A_2(y)E^2(y)+A_3(y)E^3(y)) \nonumber\\
&&\qquad\qquad\qquad+
  (E^3(y)E^2(y)'-E^2(y)E^3(y)')
  (E^2(x)A_3(y)-E^3(x)A_2(y))\bigr)\partial_x\delta(x,y)\nonumber\\
&&\qquad+ E^1(x)E^1(y) \bigl(2A_1(y)
  (E^2(x)A_3(y)-E^3(x)A_2(y)) \nonumber\\
&&\qquad\qquad\qquad
  +E^1(x)E^1(y)'(A_2(y)E^2(x)+A_3(y)E^3(x))\bigr)\partial_x\delta(x,y)
\nonumber\\
&&\qquad -2 E^1(x)E^1(y) \bigl(A_2(y)E^2(y)'+ A_3(y)E^3(y)'+ A_2(y)'E^2(y)+
  A_3(y)'E^3(y) \nonumber\\
&&\qquad\qquad\qquad - 2(A_2(y)E^2(y)+A_3(y)E^3(y))
  \frac{E^2(x)E^2(y)'+E^3(x)E^3(y)'}{E^2(y)^2+E^3(y)^2}
\Biggr)\partial_x\delta(x,y)\Biggr) \nonumber\\
&=& \frac{\gamma^2-\epsilon}{2\gamma G} \int{\rm d}x{\rm d}yN(x)M(y)
 \Biggl((\cdots)\delta(x,y)+2E^1(x)E^1(y) \nonumber\\
&&\qquad\qquad\times \bigl( A_1(y) (E^2(x)A_3(y)-E^3(x)A_2(y)) -
(A_2(y)'E^2(y)+ A_3(y)'E^3(y))\bigr) \partial_x\delta(x,y)\nonumber\\
&=& (\gamma^2-\epsilon) \left(D[(E^1)^2N'M]+
  G[A_1(E^1)^2N'M]\right)\nonumber\\
&& - \frac{\gamma^2-\epsilon}{2\gamma G} \int{\rm d}x N'M E^1(E^1)'
(A_2E^2+A_3E^3)\,. 
\end{eqnarray}

This result provides the diffeomorphism constraint as well as a term which
cancels the previous non-constraint contribution in (\ref{brack1}), but only
if $\dot{f}=1$. Therefore, if the Lorentzian contribution is included, no
modification of the classical $A_2^2+A_3^2-1$ is allowed. The final bracket
now equals
\begin{eqnarray}
	\{H[\underset{\sim}{N}],H[\underset{\sim}{M}]\}&=& \{H^{\rm
    E}[\underset{\sim}{N}],H^{\rm E}[\underset{\sim}{M}\} + \{H^{\rm
    E}[\underset{\sim}{N}],H^{\rm L}[\underset{\sim}{M}\}- \{H^{\rm
    E}[\underset{\sim}{M}],H^{\rm L}[\underset{\sim}{N}\}\nonumber\\
  &=& \gamma^2 D[(E^1)^2(\underset{\sim}{N}'\underset{\sim}{M}-
  \underset{\sim}{N}\underset{\sim}{M}')]  + \gamma^2 G[A_1
  (E^1)^2(\underset{\sim}{N}'\underset{\sim}{M}- 
  \underset{\sim}{N}\underset{\sim}{M}')]\nonumber\\
&& - (\gamma^2-\epsilon)
  G[E^1 (A_2E^2+A_3E^3)(1-2\dot{f})(\underset{\sim}{N}'\underset{\sim}{M}- 
  \underset{\sim}{N}\underset{\sim}{M}')]\nonumber\\
&& -(\gamma^2-\epsilon)
  \left(D[(E^1)^2(\underset{\sim}{N}'\underset{\sim}{M}- 
  \underset{\sim}{N}\underset{\sim}{M}')]+ G[A_1
  (E^1)^2(\underset{\sim}{N}'\underset{\sim}{M}- 
  \underset{\sim}{N}\underset{\sim}{M}')]\right) \nonumber\\
  &=& \epsilon \left(D[(E^1)^2(\underset{\sim}{N}'\underset{\sim}{M}- 
  \underset{\sim}{N}\underset{\sim}{M}')]+ G[A_1
  (E^1)^2(\underset{\sim}{N}'\underset{\sim}{M}- 
  \underset{\sim}{N}\underset{\sim}{M}')]\right) \nonumber\\
&&+ (\gamma^2-\epsilon)
  G[E^1 (A_2E^2+A_3E^3)(\underset{\sim}{N}'\underset{\sim}{M}- 
  \underset{\sim}{N}\underset{\sim}{M}')]\nonumber\\
  &\approx& -\epsilon
  D[(E^1)^2(\underset{\sim}{N}\underset{\sim}{M}'-  
  \underset{\sim}{N}'\underset{\sim}{M})]\,,
\end{eqnarray}
using $\dot{f}=1$ in the last step because the bracket would not be closed
otherwise. 
(Note that $\{H^{\rm L}[\underset{\sim}{N}],H^{\rm
L}[\underset{\sim}{M}]\}=0$, which can most easily be seen if one uses the
canonical variables $K_i$ and $E^i$, of which no spatial derivatives appear in
the Lorentzian contribution.)

\section{Connection variables in a canonical effective field theory}

We have seen a crucial difference between gravitational theories governed by
the Euclidean Hamiltonian constraint $H^{\rm E}$ and the full $H^{\rm
  E}+H^{\rm L}$, respectively. Formally, the reason is the difference in
derivative structures implied by the spin-connection terms in $H^{\rm L}$:
While $H^{\rm E}$ contains derivatives only of the spatial connection, $H^{\rm
  L}$ also contributes spatial derivatives of the triad. As a consequence, the
two versions allow different modifications while maintaining closed brackets. 

Derivative structures are best dealt with in a setting of effective field
theory, in which one formulates generic theories by selecting the basic fields
and the maximum order of derivatives to which they contribute, as well as
relevant symmetries. For our purposes, we need an adaptation of the usual
arguments to a canonical formulation, in which some derivatives may not be
explicit because they appear only if some of the canonical equations are used,
mainly in the relationship between momenta and ``velocities.'' 

In order to determine the correct derivative orders in a canonical theory, we
must first choose which of the basic fields should play the role of
configuration variables and therefore are considered free of time
derivatives. We are looking for a canonical theory of triads, which will
correspond to a space-time metric or triad theory, and therefore choose as our
basic fields a densitized spatial triad with momenta. The latter may be given
in terms of a connection or extrinsic curvature. The derivative order depends
on the quantum effects we wish to include. For now, we will analyze the
classical setting and therefore consider up to second-order derivatives of the
fields. Symmetries are implemented by the requirement that the constraint
brackets be closed, and in the classical case amount to
hypersurface-deformation brackets.

\subsection{Basic strategy}

In our explicit calculations of generic terms, we again follow the conventions
of section 2.2 and set $\gamma=1$ for simplicity.  For our effective
Hamiltonian, we choose to allow up to second-order in derivatives of
densitized triads. Since the conjugate momenta are of the form $A\sim \partial
E$, using the equations of motion for $\dot{E}$, we have the following general
form of the Hamiltonian constraint $H[\underset{\sim}{N}]=(2 G)^{-1}\int
{\rm d}x \underset{\sim}{N}(x)\mathcal{H}(x)$ with
\begin{equation}
	\mathcal{H}=
	\alpha^i(E^j,\partial E^j)A_i+\beta^{ij}(E^k)A_{ij}+
	\gamma^i(E)\partial A_i+Q(E,\partial E,\partial^2 E)\,, 
	\label{HGen}
\end{equation}
where we have introduced the notation $\partial\equiv \partial/\partial x$,
$A_{ij\cdots k}=A_i A_j\cdots A_k$ and $E^{ij\cdots k}=E^iE^j\cdots E^k$.  We
can already observe some preliminary restrictions on the coefficients
$\alpha^i(E,\partial E)$ and $Q(E,\partial E,\partial E\partial E, \partial^2
E)$. Both coefficients are initially allowed to depend on $\partial E^i$ and
$\partial^2 E^i$. But since we only allow up to second-order derivatives in
the Hamiltonian constraint, the dependence cannot be arbitrary. Specifically,
we have
\[
	\begin{cases}
		&\alpha^i=\bar{\alpha}^i(E)+\alpha^i_j(E)\partial E^j\\
		&Q=\bar{Q}(E)+a_i(E)\partial E^i+b_{ij}(E)\partial E^i\partial
                E^j 
		+c_i(E)\partial^2 E^i\,.
	\end{cases}
\]

We want the Hamiltonian density $\mathcal{H}$ to respect the classical
symmetries,
\begin{equation}
\begin{cases}
	\{\mathcal{H}(x),\mathcal{G}(y)\}&=0\\
	\{\mathcal{H}(x),\mathcal{D}(y)\}&=2
        G(\partial\mathcal{H}(x)\delta_{xy} 
	+2\mathcal{H}(x)\delta_{xy}')\\
	\{\mathcal{H}(x),\mathcal{H}(y)\}&
\approx-2 G(\partial(E^{11}\mathcal{D}(x))\delta_{xy}
	+2E^{11}\mathcal{D}(x)\delta_{xy}')\,,
\end{cases}
\end{equation}
where $G[\Lambda]=(2 G)^{-1}\int {\rm d}x \Lambda (x)\mathcal{G}(x)$ 
and $D[N]=(2 G)^{-1}\int {\rm d}x N(x)\mathcal{D}(x)$ 
are the diffeomorphism and
Gauss constraints, respectively. We have introduced the shorthand notation
$\delta_{xy}':=\partial_x\delta(x-y)$, and $\approx$ means ``equal'' when
setting $\mathcal{G}=0$ in the final step of calculation. These symmetries
will impose restrictions on the coefficients $\alpha_i,\beta^{ij},\gamma^i,Q$
in (\ref{HGen}), telling us what a generic Hamiltonian constraint looks like.

\subsection{Brackets}

The first bracket, $\{\mathcal{H},\mathcal{G}\}$, represents the restriction
to gauge-invariant terms for any allowed $\mathcal{H}$. Inserting
(\ref{HGen}), we have
\begin{eqnarray*}
		\{\mathcal{H}(x),\mathcal{G}(y)\}
		&=& 2 G\int {\rm d}z [(\alpha^1+2\beta^{1j}A_j)\delta_{xz}
		+\gamma^1\delta_{xz}'](x)\delta_{yz}'\\
		&&\qquad
                +[(\alpha^2+2\beta^{2j}A_j)\delta_{xz}+\gamma^2\delta'_{xz}](x) 
		(-A_3(y)\delta_{yz})\\
		&&\qquad
-[(\delta_{xz}\partial_2+\delta'_{xz}\partial_2)(\alpha^2)A_i
			+(\delta_{xz}\partial_2+\delta_{xz}'\partial_{2'}
			+\delta_{xz}''\partial_{2''})Q\\
			&&\qquad+\delta_{xz}\partial_2\beta^{ij}A_{ij}+
		\delta_{xz}\partial_2\gamma^i\partial A_i](x) E^3(y)\delta_{yz}\\
		&&\qquad
+[(\alpha^3+2\beta^{3j}A_j)\delta_{xz}+\gamma^3\delta'_{xz}](x)
		A_2(y)\delta_{yz})\\
		&&\qquad
-[(\delta_{xz}\partial_3+\delta'_{xz}\partial_{3'})(\alpha^3)A_i
			+(\delta_{xz}\partial_3+\delta_{xz}'\partial_{3'}
			+\delta_{xz}''\partial_{3''})Q\\
			&&\qquad+\delta_{xz}\partial_3\beta^{ij}A_{ij}+
		\delta_{xz}\partial_3\gamma^i\partial
                A_i](x)(-E^2(y)\delta_{yz})\\ 
		&=&0\,,
\end{eqnarray*}
where we have introduced further shorthand notation $\partial_i :=
\partial/\partial E^i$ and $\partial_{i'} :=
\partial/\partial(\partial_x E^i)$. To make the right-hand side of the
equation vanish, we need several cancellations. We can do this by first making
all functions depend on $x$ using delta functions and integrating over
$z$. Then we group terms with the same dependence on $A_i$ and derivatives of
$\delta_{xy}$ together and demand that each grouping vanish by itself.
(Different order of derivatives on $\delta$ may be dependent, for instance in
$\delta_{yx}'A(x)=A(y)\delta_{yx}'+\partial_y A(y)\delta_{yx}$. Therefore,
some $\delta'$ can produce terms that group with a $\delta$.)  This procedure
produces several dozens of partial differential equations which we will list
later along with those from the $\{\mathcal{H},\mathcal{D}\}$ bracket.

Inserting our form of $\mathcal{H}$ into the $\mathcal{H}$-$\mathcal{D}$
bracket, we obtain
\begin{eqnarray*}
		\{\mathcal{H}(x),\mathcal{D}(y)\}&=&2 G\int {\rm d}z 
		[\delta_{xz}(\alpha^1+2\beta^{1j}A_j)+\gamma^1 \delta'_{xz}](x)	
		(-A_1(y)\delta'_{yz})\\
		&&\qquad
-[(\delta_{xz}\partial_1+\delta_{xz}'\partial_{1'})(\alpha^i)A_i
			+\delta_{xz}\partial_1\beta^{ij}A_{ij} \\
			&&\qquad
+\delta_{xz}\partial_1\gamma^i\partial A_i+(\delta_{xz}\partial_1+
	\delta_{xz}'\partial_{1'}+\delta_{xz}''\partial_{1''})(Q)](x)
	(-\partial E^1(y)\delta_{yz})\\
	&&\qquad+[\delta_{xz}(\alpha^2+2\beta^{2j}A_j)+\gamma^2\delta_{xz}'](x)
	(\partial A_{2}(y)\delta_{yz})\\
	&&\qquad-[(\delta_{xz}\partial_2+\delta_{xz}'\partial_{2'})(\alpha^i)A_i
		+\delta_{xz}\partial_2\beta^{ij}A_{ij} \\
		&&\qquad+\delta_{xz}\partial_2\gamma^i\partial
                A_i+(\delta_{xz}\partial_2+ 
	\delta_{xz}'\partial_{2'}+\delta_{xz}''\partial_{2''})(Q)](x)
	( E^2(y)\delta'_{yz})\\
	&&\qquad+[\delta_{xz}(\alpha^3+2\beta^{3j}A_j)+\gamma^3\delta_{xz}'](x)
	(\partial A_{3}(y)\delta_{yz})\\
	&&\qquad-[(\delta_{xz}\partial_3+\delta_{xz}'\partial_{3'})(\alpha^i)A_i
		+\delta_{xz}\partial_3\beta^{ij}A_{ij} \\
		&&\qquad+\delta_{xz}\partial_3\gamma^i\partial
                A_i+(\delta_{xz}\partial_3+ 
	\delta_{xz}'\partial_{3'}+\delta_{xz}''\partial_{3''})(Q)](x)
	( E^3(y)\delta'_{yz})\\
	&=&2 G(\partial_x\mathcal{H}(x)\delta_{xy}
	+2\mathcal{H}(x)\delta_{xy}')\,.
\end{eqnarray*}
Similarly to how we dealt with the condition of gauge invariance, we first
integrate over $z$ to make all functions depend on $x$, then match term by
term with the right-hand side, expanded in $A_i$ and derivatives of
$\delta_{xy}$. Again, we obtain a few dozen partial differential equations.

We next list the partial differential equations that the coefficients of terms
in $\mathcal{H}$ have to obey. These equations will completely determine the
dependence on $E^2$ and $E^3$, leaving free functions of $E^1$ which the
$\mathcal{H}$-$\mathcal{H}$ bracket will further restrict. These conditions
then determine possible modifications of the classical $\mathcal{H}_{\rm
  cl}$. In the following equations, we use the differential operators
$\hat{D}:= E^2\partial_2+E^3\partial_3$ and $\hat{C}:=
E^2\partial_3-E^3\partial_2$.

\subsubsection{The ${\cal H}$-${\cal G}$ bracket}

For $\beta^{ij}$ and $\gamma^i$ we have
\begin{equation}
\left\{
	\begin{aligned}
		\hat{C}\beta^{11}=&0\\
                \hat{C}\beta^{12}=&-\beta^{13}\\
		\hat{C}\beta^{13}=&\beta^{12}
	\end{aligned}
	\right.
\left\{	
	\begin{aligned}
		\hat{C}\beta^{22}&=-2\beta^{23}\\
		\hat{C}\beta^{33}&=2\beta^{23}\\
		\hat{C}\beta^{23}&=\beta^{22}-\beta^{33}
	\end{aligned}
	\right.
\left\{
		\begin{aligned}
			\hat{C}\gamma^1&=0\\
			\hat{C}\gamma^2&=-\gamma^3\\
			\hat{C}\gamma^3&=\gamma^2
		\end{aligned}
		\right.
		\label{pde-hg0}
	\end{equation}
For $\alpha^i$ we have
\begin{equation}
\left\{
	\begin{aligned}
		\hat{C}\bar{\alpha}^1&=0\\
		\hat{C}\bar{\alpha}^2&=-\bar{\alpha}^3\\
		\hat{C}\bar{\alpha}^3&=\bar{\alpha}^2
	\end{aligned}
	\right.
	\left\{
		\begin{aligned}
			\hat{C}\alpha^1_1&=0\\
			\hat{C}\alpha^2_1&=-\alpha^3_1\\
			\hat{C}\alpha^3_1&=\alpha^2_1
		\end{aligned}
		\right.
	\left\{
		\begin{aligned}
			\hat{C}\alpha^1_2&=-\alpha^1_3\\
			\hat{C}\alpha^2_2&=-\alpha^3_2-\alpha^2_3\\
			\hat{C}\alpha^3_2&=\alpha^2_2-\alpha^3_3
		\end{aligned}
		\right.
		\left\{
			\begin{aligned}
				\hat{C}\alpha^1_3&=\alpha^1_2\\	
				\hat{C}\alpha^2_3&=\alpha^2_2-\alpha^3_3\\
				\hat{C}\alpha^3_3&=\alpha^3_2+\alpha^2_3
			\end{aligned}
			\right.
			\label{pde-hg1}
		\end{equation}
For $Q$ we have
\begin{equation}
\hat{C}\bar{Q}=0
\label{pde-hg2}
\end{equation}
\begin{equation}
\left\{
	\begin{aligned}
		\hat{C}a_1&=0\\
		\hat{C}a_2&=-a_3\\
		\hat{C}a_3&=a_2
	\end{aligned}
	\right.
\left\{
	\begin{aligned}
		\hat{C}b_{11}&=0\\
		\hat{C}b_{12}&=-b_{13}\\
		\hat{C}b_{13}&=b_{12}
	\end{aligned}
	\right.
\left\{
	\begin{aligned}
		\hat{C}b_{22}&=-2b_{32}\\
		\hat{C}b_{33}&=2b_{32}\\
		\hat{C}b_{23}&=b_{22}-b_{33}
	\end{aligned}
	\right.
\left\{
	\begin{aligned}
		\hat{C}c_1&=0\\
		\hat{C}c_2&=-c_3\\
		\hat{C}c_3&=c_2
	\end{aligned}
	\right.
	\label{pde-hg3}
\end{equation}
The remaining equations mix different coefficients:
\begin{equation}
\left\{
	\begin{aligned}
		&E^2a_3-E^3a_2=\bar{\alpha}^1\\
		&(-\alpha^1_j+2E^2b_{3j}-2E^3b_{2j})\partial E^j=-2(\partial E^2c_3
		-\partial E^3c_2 )\\
		&E^2c_3-E^3c_2=\gamma^1
	\end{aligned}
	\right.
\left\{
	\begin{aligned}
		&E^2\alpha^1_3-E^3\alpha^1_2=2\beta^{11}\\
		&E^2\alpha^2_3-E^3\alpha^2_2=2\beta^{12}-\gamma^3\\
		&E^2\alpha^3_3-E^3\alpha^3_2=2\beta^{13}+\gamma^2
	\end{aligned}
	\right.
	\label{pde-hg4}
\end{equation}

\subsubsection{The $\mathcal{H}$-$\mathcal{D}$ bracket}

For $\beta^{ij}$ and $\gamma^i$ we have
\begin{equation}
\left\{
	\begin{aligned}
		\hat{D}\beta^{11}&=0\\
		\hat{D}\beta^{12}&=\beta^{12}\\
		\hat{D}\beta^{13}&=\beta^{13}
	\end{aligned}
	\right.
\left\{
	\begin{aligned}
		\hat{D}\beta^{22}&=2\beta^{22}\\
		\hat{D}\beta^{33}&=2\beta^{33}\\
		\hat{D}\beta^{23}&=2\beta^{23}
	\end{aligned}
	\right.
\left\{
	\begin{aligned}
		\hat{D}\gamma^1&=0\\
		\hat{D}\gamma^2&=\gamma^2\\
		\hat{D}\gamma^3&=\gamma^3
	\end{aligned}
	\right.
	\label{pde-hd1}
\end{equation}
For $\alpha^i$ we have
\begin{equation}
\left\{
	\begin{aligned}
		\hat{D}\bar{\alpha}^1&=\bar{\alpha}^1\\
		\hat{D}\bar{\alpha}^2&=2\bar{\alpha}^2\\
		\hat{D}\bar{\alpha}^3&=2\bar{\alpha}^3
	\end{aligned}
	\right.
\left\{
	\begin{aligned}
		\hat{D}\alpha^1_1&=0\\
		\hat{D}\alpha^2_1&=\alpha^2_1\\
		\hat{D}\alpha^3_1&=\alpha^3_1
	\end{aligned}
	\right.
\left\{
	\begin{aligned}
		\hat{D}\alpha^1_2&=-\alpha^1_2\\
		\hat{D}\alpha^2_2&=0\\
		\hat{D}\alpha^3_2&=0
	\end{aligned}
	\right.
\left\{
	\begin{aligned}
		\hat{D}\alpha^1_3&=-\alpha^1_3\\
		\hat{D}\alpha^2_3&=0\\
		\hat{D}\alpha^3_3&=0
	\end{aligned}
	\right.
\left\{
	\begin{aligned}
E^2\alpha^2_2+E^2\alpha^2_3&=0\\
E^2\alpha^3_2+E^3\alpha^3_3&=0
	\end{aligned}
	\right.
	\label{pde-hd2}
\end{equation}
For $Q$ we have
\begin{equation}
\left\{
	\begin{aligned}
		&\hat{D}\bar{Q}=2\bar{Q}\\
		&E^2c_2+E^3c_3=0\\
		&E^2a_2+E^3a_3=0
	\end{aligned}
	\right.
\left\{
	\begin{aligned}
		&c_1+2(b_{12}E^2+b_{13}E^3)=0\\
		&3c_2+2(b_{22}E^2+b_{23}E^3)=0\\
		&3c_3+2(b_{32}E^2+b_{33}E^3)=0
	\end{aligned}
	\right.
	\label{pde-hd3}
\end{equation}
\begin{equation}
\left\{
	\begin{aligned}
		\hat{D}c_1&=0\\
		\hat{D}c_2&=-c_2\\
		\hat{D}c_3&=-c_3
	\end{aligned}
	\right.
\left\{
	\begin{aligned}
		\hat{D}a_1&=a_1\\
		\hat{D}a_2&=0\\
		\hat{D}a_3&=0
	\end{aligned}
	\right.
\left\{
	\begin{aligned}
		\hat{D}b_{11}&=0\\
		\hat{D}b_{12}&=-b_{12}\\
		\hat{D}b_{13}&=-b_{13}
	\end{aligned}
	\right.
\left\{
	\begin{aligned}
		\hat{D}b_{22}&=-2b_{22}\\
		\hat{D}b_{33}&=-2b_{33}\\
		\hat{D}b_{23}&=-2b_{23}
	\end{aligned}
	\right.
	\label{pde-hd4}
\end{equation}
One equation mixes different coefficients:
\begin{equation}
E^2\alpha^1_2+E^3\alpha^1_3=-\gamma^1\,.
\label{pde-hd5}
\end{equation}

\subsubsection{The $\mathcal{H}$-$\mathcal{H}$ bracket}

Matching term by term for $\mathcal{H}$-$\mathcal{H}$ is quite tedious, mainly
because the classical bracket $\{\mathcal{H},\mathcal{H}\}$ is fully
determined only after setting $\mathcal{G}=0$. For example, if there is a term
$f(\alpha,\beta,\gamma,Q) \partial E^1$ on the left-hand side of
$\{\mathcal{H}(x),\mathcal{H}(y)\}
\approx-2 G(E^{11}\partial_x\mathcal{D}(x)\delta_{xy}
+2E^{11}\mathcal{D}(x)\delta_{xy}')$ which is not on the right hand side, do
we demand $f(\alpha,\beta,\gamma,Q)=0$ or do we demand
$f(\alpha,\beta,\gamma,Q)\propto \mathcal{G}$ or $\partial \mathcal{G}$, or
does $f(\alpha,\beta,\gamma,Q)\partial E^1$ combine with possible
$f(\alpha,\beta,\alpha,Q)(-E^2A_3+E^3E_2)$ terms to become something
proportional to $\mathcal{G}$?  There are about $10^2$ terms on the left-hand
side of the $\mathcal{H}$-$\mathcal{H}$ bracket, each of which has several
possibilities of respecting the symmetry (in the form of second-order
polynomial equations of $\alpha,\beta,\gamma,Q$). It is therefore necessary to
check whether these $(10^2)^{n},n\sim 10^0$ possibilities are consistent with
one another, rendering our current strategy impractical.  Luckily, we
can use an alternative strategy to find a subset of the most generic
Hamiltonian by adding ``semi-symmetric Gaussian'' terms to the classical
Hamiltonian constraint.

\subsection{Real vs.\ self-dual variables}

We define a \textit{semi-symmetric} term to be any term in a generic
Hamiltonian constraint that is allowed by the $\{H,D\}$ and $\{H,G\}$
brackets. These terms are solutions to our previous partial differential
equations \eqref{pde-hg0}-\eqref{pde-hd5}. We define a \textit{Gaussian} term to be any term that is a
polynomial of $\mathcal{G}$ and $\partial^n\mathcal{G}$, with coefficients
denoted collectively as $C(E)$, which may depend on densitized triads and its derivatives.
Namely, for a semi-symmetric Gaussian term 
$g(x):=g[\mathcal{G}(x),\partial^n\mathcal{G}(x),C(E(x))]$ we demand 
\begin{equation}
\begin{cases}
\{g(x),\mathcal{G}(y)\}&=0\\
\{g(x),\mathcal{D}(y)\}&=2
G(\partial g(x)\delta_{xy} 
+2g(x)\delta_{xy}')\,,
\end{cases}
\end{equation}
Any semi-symmetric Gaussian term, $g[\mathcal{G},\partial^n\mathcal{G}, C(E)]$,
that we add to the classical Hamiltonian constraint $\mathcal{H}_{cl}$ is
guaranteed to respect all our symmetries as shown below.

Suppose we add one semi-symmetric Gaussian term
$g[\mathcal{G},\partial^n\mathcal{G},C(E)]$ to the classical Hamiltonian
constraint ${\cal H}_{\rm cl}$
\begin{equation}
	H[\underset{\sim}{N}]=\frac{1}{2 G}\int {\rm d}x
        \underset{\sim}{N}(x)( 
        \mathcal{H}_{\rm cl}+g) \,. 
	\label{}
\end{equation}
Since $\mathcal{H}_{\rm cl}$ respects all symmetries by definition and $g$ is
built out of semi-symmetric Gaussian terms,
\begin{equation}
	\{H[\underset{\sim}{N}],G[M]\}=0
	\label{}
\end{equation}
is trivial. Similarly, the $H$-$D$ bracket is satisfied:
\begin{equation}
	\begin{aligned}
		\{H[\underset{\sim}{N}],D[M]\}&=
		\frac{1}{4 G^2}\int {\rm d}x {\rm d}y
                \underset{\sim}{N}(x)M(y)(\{\mathcal{H}_{\rm cl},\mathcal{D}\} 
		+\{g,\mathcal{D}\})\\
		&=\frac{1}{2 G}\int {\rm d}x {\rm d}y
                \underset{\sim}{N}(x)M(y)(\partial_x\mathcal{H}_{\rm cl}(x) 
		\delta_{xy}+2\mathcal{H}_{\rm cl}(x)\delta_{xy}'+\partial_x
                g(x)\delta_{xy}+ 
		2g(x)\delta_{xy}')\\
		&=\frac{1}{2 G}\int {\rm d}x{\rm
                  d}y\underset{\sim}{N}(x)M(y) 
		(\partial_x \mathcal{H}(x)\delta_{xy}+2\mathcal{H}(x)\delta_{xy}')
		=-H[M\underset{\sim}{N}'-M'\underset{\sim}{N}]
	\end{aligned}
\end{equation}
because $g$ is built out of semi-symmetric Gaussian terms.  The
$H[\underset{\sim}{N}]$-$H[\underset{\sim}{M}]$ bracket then has additional
terms compared with the classical case, given by $\{{\cal H}_{\rm cl},g\}$ and
$\{g,g\}$. Both terms are of the form $\{f,g\}$ with some semi-symmetric $f$,
and share the property that $\int{\rm d}x{\rm d}yN(x)M(y) \{f(x),g(y)\}$
vanishes when ${\cal G}=0$: In
\begin{eqnarray}
  &&\int {\rm d}x{\rm d}y
  N(x)M(y)\{f(x),g[\mathcal{G}(y),\partial^n\mathcal{G}(y), C(E)]\}\nonumber\\ 
  &=&
  \int {\rm d}x{\rm d}y N(x)M(y)\left(\{f(x),\mathcal{G}(y)\}\frac{\partial
    g}{\partial\mathcal{G}}(y) 
  +\{f(x),\partial_y^n\mathcal{G}(y)\}
  \frac{\partial g}{\partial(\partial^n_y\mathcal{G})}(y)\right.\nonumber\\
  &&+\left.\{f(x),C(E)\}\frac{\partial g}{\partial C(E)}\right)\nonumber\\
  &=&\int {\rm d}x{\rm d}y N(x)M(y)\left(\{f(x),\mathcal{G}(y)\}\frac{\partial
    g}{\partial\mathcal{G}}(y)+ 
  \{f(x),C(E)\}\frac{\partial g}{\partial C(E)}\right)\nonumber\\
  &&+\int {\rm d}x{\rm d}y N(x) (-\partial_y)^n
  \left(M(y)\frac{\partial g}{\partial(\partial^n_y\mathcal{G})}(y)\right)
  \{f(x),\mathcal{G}(y)\}\,,
\end{eqnarray}
the first and last term vanish because $f$ is semi-symmetric, while $\partial
g/\partial C(E)\approx 0$ because $C(E)$, by definition, represents
coefficients in $g$ of the Gauss constraint or its spatial derivatives.

With this result, we confirm that
\begin{eqnarray}
	\{H[\underset{\sim}{N}],H[\underset{\sim}{M}]\}&=& \frac{1}{4
          G^2} 
		\int {\rm d}x{\rm d}y\underset{\sim}{N}(x)\underset{\sim}{M}(y)
		(
		\{\mathcal{H}_{\rm cl}(x),\mathcal{H}_{\rm cl}(y)\}\nonumber\\
		&&+\{g[\mathcal{G}(x),\partial^n\mathcal{G}(x),C(E)],
		g[\mathcal{G}(y),\partial^n\mathcal{G}(y),C(E)]\}\nonumber\\
		&&+\{\mathcal{H}_{\rm
                  cl}(x),g[\mathcal{G}(y),\partial^n\mathcal{G}(y),C(E)]\}
 \nonumber
                \\  
		&&
+\{g[\mathcal{G}(x),\partial^n\mathcal{G}(x),C(E)],\mathcal{H}_{\rm
                  cl}(y)\} 
		)\nonumber\\
		&\approx& \frac{1}{4 G^2}\int {\rm d}x{\rm
                  d}y\underset{\sim}{N}(x)\underset{\sim}{M}(y) 
		\{\mathcal{H}_{\rm cl}(x),\mathcal{H}_{\rm cl}(y)\}\nonumber\\
\end{eqnarray}
obeys the classical brackets for any semi-symmetric $g$.  Thus,
semi-symmetric Gaussian terms indeed preserve all symmetries.

When written in real variables, the classical Hamiltonian constraint contains
a term with second-order derivative of $E^1\sim E^{x}$, given by
$2\partial\Gamma_{\phi}E^x=-\partial(\partial E^{x}/(E^{\varphi}))E^x$.  But
when using self-dual variables, there are no second-order derivative of
triads. As already mentioned, this discrepancy is caused by the fact that
$\mathcal{G}\approx 0$ is already solved in the real variable case. Indeed,
using semi-symmetric terms (see appendix \ref{appendix-pdesol}) 
for constructing modifications we
have the following allowed terms when using self-dual variables
\begin{eqnarray}
		\mathcal{H}_{2}(A,E)&=&\mathcal{H}_{\rm
                  cl}(A,E)+c_1(E^1)\left(\partial \mathcal{G} 
		-\frac{1}{2}\frac{\partial((E^{\varphi})^2)}{(E^{\varphi})^2}
 \mathcal{G}\right)\nonumber\\  
   && +\partial E^1[b_{11}(E^1)\partial
        E^1+\tilde{C}_{\alpha^2_1}(E^1)(E^3A_2-E^2A_3)] \,,
	\label{min-k}
\end{eqnarray}
where $\partial\mathcal{G}\sim \partial^2 E^1$ provides the second-order
derivative. Note that the second semi-symmetric term (proportional to
$\partial E^1$) becomes a semi-symmetric Gaussian term if we pick
$b_{11}=\frac{1}{2}\tilde{C}_{\alpha^2_1}$.

Substituting $A_i=\gamma K_i+\Gamma_i$, $c_1=E^1,b_{11}=\frac{1}{2}\tilde{C}_{\alpha^2_1}
=1/2$ in the classical Hamiltonian
constraint and de-densitizing, we obtain
\begin{equation}
	\mathcal{H}_{2}(K,E)=|E^x|^{-1/2}\left(K_{\varphi}^2
        E^{\varphi}+2K_{\varphi}K_x E^x- 
	\left(1-\left(\frac{\partial
          E^{x}}{2E^{\varphi}}\right)^2\right)E^{\varphi}+
 \frac{E^x\partial^2E^x}{E^{\varphi}}-  
\frac{E^x\partial E^x\partial E^{\varphi}}{(E^{\varphi})^2}\right) \,,
	\label{min-ham}
\end{equation}
where we used the Gauss constraint in real variables.  This result matches the
standard classical Hamiltonian constraint in real variables. Thus, including
semi-symmetric Gaussian terms in the quadratic constraint, it is equivalent to
the classical one written in real variables.

Revisiting the setting of the previous section, it follows that a further
restriction of our $\mathcal{H}$ to be only quadratic in densitized triads
implies that all allowed modifications to the classical $\mathcal{H}_{\rm cl}$
are in the form of semi-symmetric Gaussian terms:
\begin{eqnarray}
	\mathcal{H}_{\rm quad}&=&C_1(\partial A_3 E^{21}-\partial A_2
        E^{31}+A_{12}E^{12} 
	+A_{13}E^{13})+C_2\left(A_{22}+A_{33}+\frac{C_3}{C_2}\right)
        (E^{22}+E^{33})\nonumber \\
	&&+C_4\partial E^1\mathcal{G}+C_5(A_2E^2+A_3E^3)\mathcal{G}\,.
\end{eqnarray}
The first two terms are present in $\mathcal{H}_{\rm cl}$ while the last two
are new semi-symmetric Gaussian terms and all $C_i$ are constants. However,
the complexity of the general equations makes it difficult to show that all
possible modifications to the Hamiltonian constraint up to second order in
derivatives can be constructed from semi-symmetric Gaussian terms.

\section{Eliminating the Gauss constraint}

Our analysis of gravitational theories in a setting of effective field theory
has highlighted the role of the Gauss constraint, which implies that the
hypersurface-deformation generators are not uniquely defined. Since the Gauss
constraint contains a spatial derivative, and spatial derivatives of this
constraint can also be added to the hypersurface-deformation generators, the
derivative structure and therefore the possibility of modifications is
ambiguous as long as the Gauss constraint remains unsolved. We will therefore
now solve the Gauss constraint explicitly and analyze the resulting
hypersurface-deformation generators and their brackets.

\subsection{Gauge-invariant variables}

We begin with the classical constraint
\begin{eqnarray}
\label{ham1}
H[N] &=& \int {\rm d}x
\frac{N}{\sqrt{E^1((E^2)^2+(E^3)^2)}}\left(2E^1(E^2A_3'-E^3A_2')\right.\\ 
&&+2A_1E^1(A_2E^2+A_3E^3)+(A_2^2+A_3^2-1)((E^2)^2+(E^3)^2)\nonumber\\
&&+\left.(\epsilon-\gamma^2)
(2K_1E^1(K_2E^2+K_3E^3)+(K_2^2+K_3^2)((E^2)^2+(E^3)^2)\right) \nonumber
\end{eqnarray}
in which the lapse function no longer has a density weight. The next few
transformations closely follow the derivations given in \cite{SphSymmHam}, but
are presented here in a different form using vector notation.

The pairs $(E^2,E^3)$ and $(A_2,A_3)$ (as well as $(K_2,K_3)$) transform under
the defining representation of ${\rm SO}(2)$ with respect to the Gauss
constraint. It will be convenient to arrange them in 3-vectors, such that
\begin{equation}
 \vec{E}=E^2\vec{e}_2+E^3\vec{e}_3 \quad,\quad
 \vec{A}=A_2\vec{e}_2+A_3\vec{e}_3 \quad,\quad
 \vec{K}=K_2\vec{e}_2+K_3\vec{e}_3 
\end{equation}
with standard basis vectors $\vec{e}_i$. Obvious
invariant variables are therefore
\begin{equation}
 E^{\varphi}=|\vec{E}|=\sqrt{(E^2)^2+(E^3)^2}\quad,\quad
A_{\varphi}=|\vec{A}|=\sqrt{A_2^2+A_3^2}\quad,\quad
K_{\varphi}=|\vec{K}|=\sqrt{K_2^2+K_3^2}\,.
\end{equation}
Moreover, we obtain another invariant $\alpha$ from the angle between
$\vec{E}$ and $\vec{A}$,
\begin{equation}
 \cos\alpha=\frac{\vec{E}\cdot\vec{A}}{E^{\varphi}A_{\varphi}}\,.
\end{equation}
While $E^1$ and $K_1$ are also invariant, $A_1$ has a non-trivial
transformation. A final gauge-invariant expression can be written as
$A_1+\beta'$, where 
\begin{equation}
 \cos\beta=\frac{\vec{e}_2\cdot\vec{A}}{A_{\varphi}}\,.
\end{equation}

Using our definitions of $\alpha$ and $\beta$, we can write the unit vectors
\begin{eqnarray}
 \vec{e}_A&=& \frac{\vec{A}}{A_{\varphi}}= \vec{e}_2\cos(\beta)+\vec{e}_3
 \sin(\beta)\\ 
 \vec{e}_E&=& \frac{\vec{E}}{E^{\varphi}}= \vec{e}_2\cos(\alpha+\beta)+\vec{e}_3
 \sin(\alpha+\beta)\,.
\end{eqnarray}
From the last relation one can derive the spin-connection component
$\Gamma_1=-(\alpha+\beta)'$ \cite{SphSymmHam}. Therefore,
$\gamma^{-1}(A_1+\alpha'+\beta')=K_1$ is nothing but an extrinsic-curvature
component. Since $\alpha$ and $K_1$ are gauge invariant, $A_1+\beta'$ must be
gauge invariant, as claimed above.

Moreover, computing the extrinsic curvature and spin connection for a
spherically symmetric triad \cite{SphSymmHam} shows that the angular part
$\vec{K}$ points in the same internal direction as the triad,
\begin{equation}
 \vec{e}_K=\vec{e}_E\,,
\end{equation}
while the angular part of the spin connection, $\vec{\Gamma}$, is orthogonal,
\begin{equation}
 \vec{e}_{\Gamma}=-\vec{e}_1\times\vec{e}_E\,,
\end{equation}
with coefficient
\begin{equation}
 \Gamma_{\varphi}=-\frac{(E^1)'}{2E^{\varphi}}\,;
\end{equation}
see (\ref{spinconn}).  Therefore,
\begin{equation}
 A_{\varphi}^2=|\vec{A}|^2= |\Gamma_{\varphi}\vec{e}_{\Gamma}+\gamma
 K_{\varphi}\vec{e}_K|^2= \Gamma_{\varphi}^2+\gamma^2K_{\varphi}^2\,.
\end{equation}

The term in (\ref{ham1}) containing spatial derivatives of the connection can
now be written as
\begin{equation}
 E^2A_3'-E^3A_2' = \vec{e}_1 \cdot (\vec{E}\times\vec{A}')= E^{\varphi}
 \vec{e}_1 (\vec{e}_E\times (A_{\varphi}\vec{e}_A)'=
 E^{\varphi}(-A_{\varphi}'\sin(\alpha)+ A_{\varphi}\beta'\cos(\alpha))\,.
\end{equation}
We then express connection terms through spin connection and extrinsic
curvature, using
\begin{equation}
 A_{\varphi}\sin(\alpha)= A_{\varphi} \vec{e}_A\cdot\vec{e}_{\Gamma}=
 \Gamma_{\varphi}
\end{equation}
and
\begin{equation}
 A_{\varphi}\cos(\alpha)= A_{\varphi} \vec{e}_A\cdot\vec{e}_{K}=
 \gamma K_{\varphi}\,.
\end{equation}
Therefore,
\begin{equation}
 E^2A_3'-E^3A_2' = E^{\varphi}(-(A_{\varphi}\sin(\alpha))+
 A_{\varphi}(\alpha'+\beta')\cos(\alpha)) =
 E^{\varphi}\left(-\Gamma_{\varphi}'+\gamma
   K_{\varphi}(\alpha'+\beta')\right)\,.
\end{equation}
The angles in the last term can be combined with a similar contribution from
the second term in (\ref{ham1}), which adds $A_1$ to $\alpha'+\beta'$. (In
(\ref{ham1}), $A_1$ is multiplied with
$A_2E^2+A_3E^3=\vec{A}\cdot\vec{E}=\gamma K_{\varphi}E^{\varphi}$, which does
not depend on $\Gamma_{\varphi}$ because $\vec{e}_{\Gamma}\cdot\vec{e}_E=0$.)
Since $\alpha'+\beta'=-\Gamma_1$ \cite{SphSymmHam} and $A_1-\Gamma_1=\gamma
K_1$, we have 
\begin{equation}
 E^2A_3'-E^3A_2' +A_1(A_2E^2+A_3E^3)= E^{\varphi}\left(-\Gamma_{\varphi}'+\gamma^2
   K_{\varphi}K_1\right)\,.
\end{equation}
Thus, by using variables invariant under transformations generated by the
Gauss constraint, we have been led to an expression in which all spatial
derivatives of the connection have been replaced by spatial derivatives of the
triad (through $\Gamma_{\varphi}$).

Again in \cite{SphSymmHam}, the Poisson brackets
\begin{equation}
\{ K_{\varphi}(x), E^{\varphi}(y) \} = G \delta(x,y), \quad \{ K_1(x), E^1(y) \} =
2G\delta(x,y)
\end{equation}
for the new gauge-invariant variables have been derived. If we express the
diffeomorphism and Hamiltonian constraints in these variables, we 
restrict the previous theory to the solution space of the Gauss constraint. We
obtain 
\begin{equation}
D[N^x] =\frac{1}{2G} \int {\rm d}x N^x \left(2E^{\varphi} K'_{\varphi} -
  K_1(E^1)' \right)
\end{equation}
and
\begin{equation}
\label{hamK}
H[N] = \frac{1}{2G} \int {\rm d}x \frac{N}{\sqrt{E^1}}
\left(K_{\varphi}^2E^{\varphi}(\epsilon-\gamma^2)+2\epsilon K_{\varphi} 
K_1E^1+(\Gamma_{\varphi}^2+\gamma^2K_{\varphi}^2-1)E^{\varphi}-
2E^1\Gamma'_{\varphi}\right)\,. 
\end{equation}

\subsection{Modified constraint with classical brackets}

In the Hamiltonian constraint, the two terms with $\gamma^2K_{\varphi}^2$
cancel out, showing that, for $\epsilon=-1$, we obtain the Hamiltonian
constraint as considered in \cite{SphSymmHam}. Our calculation here extends
this result to Euclidean signature, $\epsilon=1$. Since all $\gamma$-dependent
terms drop out of the final expression, it is no longer clear why
$\gamma^2=\epsilon$ should lead to different options for modified
constraints. Nevertheless, the previous distinction between
$\gamma^2=\epsilon$ and $\gamma^2\not=\epsilon$ can still be realized if we do
not cancel the $\gamma$-dependent terms in (\ref{hamK}) {\em before} we try to
modify the constraint. In particular, the previous modification, using an
arbitrary function of $f(A_2^2+A_3^2-1)$, can still be implemented in the
invariant version if we recognize the combination
$\Gamma_{\varphi}^2+\gamma^2K_{\varphi}^2-1$ as the correct substitute of
$A_2^2+A_3^2-1= A_{\varphi}^2-1$. We therefore consider the modified constraint
\begin{equation}
\label{hamKf}
H[N] = \frac{1}{2G} \int {\rm d}x
\frac{N}{\sqrt{E^1}}\left(K_{\varphi}^2E^{\varphi}(\epsilon-\gamma^2)+2\epsilon
  K_{\varphi}  K_1E^1+f(\Gamma_{\varphi}^2+\gamma^2K_{\varphi}^2-1)E^{\varphi}
-2E^1\Gamma'_{\varphi}\right)\,.
\end{equation}
Given the form of this new constraint, it is not obvious that it can lead to
closed brackets for $f$ not the identity because, compared with our previous
derivation, we now have up to second-order spatial derivatives of the triad
(through $\Gamma_{\varphi}$) instead of first-order derivatives of its momenta.

Thanks to antisymmetry of the Poisson bracket, the only terms that give
non-zero contributions to $B_{NM}:=\{ H[N], H[M] \}$ are combinations of a
term from $H[N]$ depending on one of the $K_i$ and a term from $H[M]$
depending on a (first or second order) spatial derivative of one of the $E_i$,
or vice versa.  Therefore,
\begin{eqnarray}
B_{NM} &=& \frac{1}{4G^2}\int {\rm d}x{\rm d}y
\frac{N(x)M(y)}{\sqrt{E^1(x)E^1(y)}}
\left(-(\epsilon-\gamma^2)\{ K^2_{\varphi}(x), (E^{\varphi})'
\}
\frac{E^1(y)E^1(y)'E^{\varphi}(x)}{(E^{\varphi}(y))^2}\right.\nonumber\\
&& - 2\epsilon \{K_{\varphi}(x), E^{\varphi}(y)'
\}K_1(x)\frac{E^1(x)E^1(y)E^1(y)'}{(E^{\varphi}(y))^2} \nonumber\\
&&
-2\epsilon K_{\varphi}(x)\{ K_1(x), E^1(y)'
\}
\frac{E^1(x)E^1(y)E^{\varphi}(y)'}{(E^{\varphi}(y))^2}\nonumber\\
&&- 
\{f, E^{\varphi}(y)'
\}\frac{E^{\varphi}(x)E^1(y)E^1(y)'}{(E^{\varphi}(y))^2}+2\epsilon
K_{\varphi}(x)\{K_1(x), f \}E^1(x)E^{\varphi}(y)\nonumber\\
&&
\left.+2\epsilon K_{\varphi}(x)\frac{E^1(y)}{E^{\varphi}(y)}\{ 
K_1(x), E^1(y)'' \}E^1(x)\right)-(N\leftrightarrow M)\,.
\end{eqnarray}
Integrating by parts, we obtain 
\begin{eqnarray}
B_{NM} &=& \frac{1}{4G}\int {\rm d}x
NM'\left((2(\epsilon-\gamma^2)K_{\varphi} 
\frac{(E^1)'}{E^{\varphi}} +2\epsilon
\frac{E^1}{(E^{\varphi})^2}K_1(E^1)'+4\epsilon K_{\varphi} (E^{\varphi})' 
\frac{E^1}{(E^{\varphi})^2}\right.\nonumber\\
&&\left.-4\epsilon \frac{E^1}{(E^{\varphi})^2}E^{\varphi} K'_{\varphi}
-4\epsilon K_{\varphi}\frac{E^1(E^{\varphi})'}{(E^{\varphi})^2}+\frac{\partial
  f}{\partial 
  K_{\varphi}}\frac{(E^1)'}{E^{\varphi}}-4\epsilon K_{\varphi} E^{\varphi}
\frac{\partial f}{\partial 
  (E^1)'}\right)-(N\leftrightarrow M) \nonumber\\
&=& \frac{-\epsilon}{2G}\int {\rm d}x
\frac{E^1}{(E^{\varphi})^2}(NM'-N'M)(2E^{\varphi} K'_{\varphi} 
-K_1(E^1)')\nonumber\\
&& +\frac{1}{4G}\int {\rm d}x (NM'-N'M)
\left(2(\epsilon-\gamma^2)K_{\varphi} \frac{(E^1)'}{E^{\varphi}}
+\frac{\partial f}{\partial 
  K_{\varphi}}\frac{(E^1)'}{E^{\varphi}}-4\epsilon K_{\varphi} E^{\varphi}
\frac{\partial   f}{\partial (E^1)'}\right) 
\nonumber \\
 &=& -\epsilon D\left[\frac{E^1}{(E^{\varphi})^2}(NM'-N'M)\right]\\
&&+\frac{1}{4G}\int {\rm d}x
 (NM'-N'M)\left(2(\epsilon-\gamma^2)K_{\varphi} \frac{(E^1)'}{E^{\varphi}}
 +\frac{\partial f}{\partial K_{\varphi}}\frac{(E^1)'}{E^{\varphi}}-4\epsilon
 K_{\varphi}  E^{\varphi} 
 \frac{\partial f}{\partial (E^1)'}\right) \,.\nonumber
\end{eqnarray}

For a closed bracket, therefore,
\begin{equation} \label{cond}
  2(\epsilon-\gamma^2)K_{\varphi} \frac{(E^1)'}{E^{\varphi}}+\frac{\partial
    f}{\partial 
    K_{\varphi}}\frac{(E^1)'}{E^{\varphi}}-4\epsilon K_{\varphi} E^{\varphi}
  \frac{\partial f}{\partial (E^1)'} = 
  0 \, .
\end{equation}
Since $f$ depends on $K_{\varphi}$ and $(E^1)'$ only through
$\frac{1}{4}(E^1{}')^2/(E^{\varphi})^2+\gamma^2K_{\varphi}^2-1$, the chain rule
implies that
\begin{equation}
 \frac{\partial f}{\partial  K_{\varphi}}=
 2\gamma^2K_{\varphi}\dot{f}\quad\mbox{and}\quad \frac{\partial f}{\partial
   (E^1)'}= 
 \frac{1}{2(E^{\varphi})^2} (E^1)' \dot{f}\,,
\end{equation}
and (\ref{cond}) is equivalent to
\begin{equation}
2(\epsilon-\gamma^2)K_{\varphi} \frac{(E^1)'}{E^{\varphi}}
\left(1-\dot{f}\right)=0\,.
\end{equation}
If $\gamma^2=\epsilon$, the equation holds identically for any $f$.  If
$\gamma^2\not=\epsilon$, however, $\dot{f}= 1$, and only the classical case is
allowed. The modification found in \cite{SphSymmComplex} can therefore be
found also in gauge-invariant variables, in which case the
  Hamiltonian constraint contains second-order derivatives of the triad, with
the same restriction that it is allowed only for a specific value of $\gamma$.

\subsection{Modified brackets}

A generic modification which does not require a specific value of $\gamma$ can
be obtained for the theories considered here, as has been known for some time
for real variables \cite{JR,HigherSpatial}. Since the Hamiltonian constraint
in real variables has the same form as the general spherically symmetric
constraint in gauge-invariant variables, the same modification can be
transferred also to self-dual type variables ($\gamma^2=\epsilon$) provided we
implement it at the gauge-invariant level. At the level of variables that are
not gauge invariant, this new modification (compared with
\cite{SphSymmComplex}) is possible provided we use the Gauss constraint to
reintroduce second-order derivatives of triads in the Hamiltonian constraint.

Starting with (\ref{hamK}), the new modification is derived in a way very
similar to the case of real variables, found in \cite{JR}. Nevertheless, we
reproduce the calculation of brackets here for the sake of completeness.
We modify (\ref{hamK}) to
\begin{eqnarray}
H[N] &=&\frac{1}{2G} \int {\rm d}x N(x)(E^1)^{-1/2}\Bigl( \epsilon
f_1(K_\varphi)E^\varphi + 
2\epsilon f_2(K_\varphi)E^1K_1\nonumber\\
&&+\left(\frac{(E^1{}')^2}{4(E^\varphi)^2}-1\right)E^\varphi+\frac{E^1
  (E^1)''}{E^\varphi}-\frac{E^1(E^1)'(E^{\varphi})'}{(E^\varphi)^2}\Bigr) 
\label{Hf1f2} 
\end{eqnarray}
with two functions, $f_1$ and $f_2$, that will be restricted further by the
condition of having closed brackets.  We first interpret this modification
based on arguments within canonical effective field theory. We are now
allowing for a non-quadratic dependence of the Hamiltonian constraint on
$K_{\varphi}$. If $K_{\varphi}$ is still a first-order time derivative, a
non-quadratic dependence would be non-generic unless we also allow for
higher-order spatial derivatives of the densitized triad, which we do not do
in (\ref{Hf1f2}).

However, modifying the Hamiltonian constraint in this form also modifies the
equations of motion that classically imply the first-order nature of
$K_{\varphi}$. An analysis of these modified equations of motion should then
be performed in order to reveal the derivative order of the Hamiltonian
constraint.  Schematically, we obtain the modified derivative dependence of
$K_{\varphi}$ from the equation of motion
\begin{eqnarray} \label{EOM}
	\dot{E}^1&=& 2N\sqrt{E^1}f_2(K_{\varphi})+ N^1 (E^1)'\\
\dot{E}^{\varphi} &=&   N \sqrt{E^1} K_1 \frac{{\rm d}f_2(K_{\varphi})}{{\rm
    d}K_{\varphi}} +  \frac{N E^{\varphi}}{2\sqrt{E^1}} \frac{{\rm
    d}f_1(K_{\varphi})}{{\rm d}K_{\varphi}} + (N^1 E^{\varphi})'
\end{eqnarray}
provided we can invert the function $f_2$. This can explicitly be done only in
examples, which we restrict here to the common case of
$f_1(K_{\varphi})=\sin^2(K_{\varphi})$, which implies
$f_2(K_{\varphi})=\sin(K_{\varphi})\cos(K_{\varphi})$ or
$f_2(K_{\varphi})^2=f_1(K_{\varphi})(1-f_1(K_{\varphi}))$. The latter equation
can be solved for 
\begin{equation}
 f_1(K_{\varphi})=\frac{1}{2}\left(1\pm\sqrt{1-4f_2(K_{\varphi})^2}\right)=
 f_2(K_{\varphi})^2+f_2(K_{\varphi})^4+\cdots \,. 
\end{equation}
According to (\ref{EOM}), $f_2(K_{\varphi})$ is strictly of first order in
derivatives, but $f_1(K_{\varphi})$ is not polynomial in
$f_2(K_{\varphi})$, and therefore a derivative expansion of $f_1(K_{\varphi})$
does not terminate. Similarly, 
\begin{equation}
 \frac{{\rm d}f_2(K_{\varphi})}{{\rm
  d}K_{\varphi}}=\cos(2K_{\varphi})=
1-2f_1(K_{\varphi})=\sqrt{1-f_2(K_{\varphi})^2}
\end{equation}
has a derivative expansion that does not terminate. Therefore, $K_1$ has a
non-terminating derivative expansion because $K_1 \sqrt{1-f_2(K_{\varphi})^2}$
must be of first order according to (\ref{EOM}). 

We conclude that the constraint (\ref{Hf1f2}) contains a derivative expansion
in both space and time derivatives, which can consistently be truncated at any
finite derivative order. The resulting effective theory is therefore
meaningful, but it may not be the most general one because the derivative
expansion results only from the $K$-dependent terms in (\ref{Hf1f2}) while we
have not included higher-derivative corrections of the $E$-dependent terms.
The mismatch does not violate (deformed) covariance because the constraint
brackets still close. However, unless the symmetries implied by the closed
constraints select only this specific derivative structure, the modified
theory is not generic. (It resembles Born--Infeld type theories.) Since no
other consistent modifications are known as of now, it remains unclear whether
the apparently non-generic model is selected by symmetries.

In order to confirm that the constraint brackets can be closed, we compute
\begin{eqnarray}
\label{pb}
\{ H[N], H[M]\} &=& \frac{1}{4G^2}\int {\rm d}x{\rm
  d}y \frac{N(x)M(y)}{\sqrt{E^1(x)E^1(y)}}\left(
-\epsilon\frac{E^\varphi(x)E^1(y)E^1(y)'}{(E^\varphi)^2(y)}\{ f_1(K_\varphi(x)),
E^{\varphi}(y)'\}  \right. \nonumber\\ 
&&-2\epsilon
\frac{E^1(x)E^1(y)E^1(y)'K_1(x)}{(E^\varphi)^2(y)}
\{f_2(K_{\varphi}(x),E^{\varphi}(y)'\} \nonumber\\
&&+\epsilon \frac{f_2(K_\varphi(x))E^1(x)}{2E^\varphi(y)}
\{K_1(x),(E^1(y)')^2\}\nonumber\\ 
&&+2\epsilon f_2(K_\varphi(x))E^1(x)\frac{E^1(y)}{E^\varphi(y)}
\{K_1(x), E^1(y)''\}  \nonumber\\
\label{pb2}
&&\left. -2\epsilon
  f_2(K_\varphi(x))E^1(x)\frac{E^1(y)E^{\varphi}(y)'}{E^\varphi(y)^2} \{K_1(x),
E^1(y)'\}\right)-
(N\leftrightarrow M)\,,
\end{eqnarray}
writing only terms that produce non-zero contributions. All terms are
multiplied with $\epsilon$, and therefore the possibility of modifications
does not depend on the space-time signature.

The first two lines contain Poisson brackets of $f_1(K_{\varphi})$ and
$f_2(K_{\varphi})$ and therefore lead to derivatives of the modification
functions:
\begin{equation} \label{pb3}
\frac{1}{G}\frac{E^\varphi(x)E^1(y)E^1(y)'}{(E^\varphi)^2(y)}\{ f_1(K_\varphi(x)),
E^{\varphi}(y)'\}= \frac{E^\varphi(x)E^1(y)E^1(y)'}{(E^\varphi)^2(y)}
\frac{{\rm d}f_1(K_{\varphi})}{{\rm d}K_{\varphi}} \partial_y\delta(x,y)
\end{equation}
and
\begin{equation} \label{pb4}
\frac{2}{G} \frac{E^1(x)E^1(y)E^1(y)'K_1(x)}{(E^\varphi)^2(y)}
\{ f_2(K_\varphi(x)), E^{\varphi}(y)'\}=
2 \frac{E^1(x)E^1(y)E^1(y)'K_1(x)}{(E^\varphi)^2(y)} 
\frac{{\rm d}f_2(K_\varphi)}{{\rm
    d}K_\varphi}\partial_y\delta(x,y)\,.
\end{equation}
Another derivative of $f_2(K_{\varphi})$ results from the second-order
derivative of the delta function obtained after evaluating $\{K_1,(E^1)''\}$
in the fourth line of (\ref{pb}). This contribution follows from
\begin{equation} \label{delta2}
2f_2(K_\varphi(x))\frac{E^1(x)E^1(y)}{E^\varphi(y)}\{ K_1(x), E^1(y)'' \} =
4f_2(K_\varphi(x))\frac{E^1(x)E^1(y)}{E^\varphi(y)} \partial_y^2\delta(x,y)\,.
\end{equation}

Upon integrating by parts twice in the resulting expression in (\ref{pb}), we
initially produce a term with $N(x)M(y)''$ times a delta function without
derivatives. Integrating over $y$, the delta function is eliminated and we can
integrate by parts once again to obtain a term with $N'M'$ (which cancels
out in the antisymmetric bracket) and a term with $NM'$ times the derivative
of the entire coefficient in (\ref{delta2}):
\begin{equation} \label{pb5}
 -4 \left(f_2(K_{\varphi}) \frac{(E^1)^2}{E^\varphi}\right)'
=  -4\left(\frac{{\rm d}f_2}{{\rm
    d}K_\varphi}K'_\varphi \frac{(E^1)^2}{E^\varphi}+
f_2(K_{\varphi})\left(2\frac{E^1(E^1)'}{E^{\varphi}}-
  \frac{(E^1)^2(E^{\varphi})'}{(E^{\varphi})^2}\right)\right) \,.
\end{equation}
The last term (containing $(E^{\varphi})'$) cancels out with the fifth line of
(\ref{pb}), while only half the second term cancels out with the third line of
(\ref{pb}), for any $f_2$. In order for the remaining terms to be proportional
to the diffeomorphism constraint, only expressions proportional to $K_1$ or
$K_{\varphi}'$ can remain. Therefore, the other half of the second term in
(\ref{pb5}) must cancel out with  (\ref{pb3}), which requires
\begin{equation}
 f_2(K_{\varphi})= \frac{1}{2}\frac{{\rm d}f_1(K_{\varphi})}{{\rm d}K_{\varphi}}\,.
\end{equation}
Only two terms are then left, (\ref{pb4}) and the first contribution in
(\ref{pb5}). They are both proportional to ${\rm d}f_2(K_{\varphi})/{\rm
  d}K_{\varphi}$ and combine to form the diffeomorphism constraint:
\begin{eqnarray}
\{ H[N], H[M]\} &=& -\frac{\epsilon}{2G} \int {\rm d}x
N'M\frac{E^1}{(E^\varphi)^2} \frac{{\rm 
    d}f_2}{{\rm d}K_\varphi}(2E^\varphi K'_\varphi -
K_1(E^1)')-(N\leftrightarrow M) \nonumber\\
&=& -\epsilon D\left[\frac{{\rm d}f_2(K_\varphi)}{{\rm
    d}K_\varphi}\frac{E^1}{(E^\varphi)^2}(NM'-N'M)\right] \, .
\end{eqnarray}

This modification, following \cite{JR,HigherSpatial}, differs from the
modification of \cite{SphSymmComplex} in that it modifies not only the
constraints but also their brackets (while the latter remain closed). It
therefore implies a new, non-classical space-time structure
\cite{Normal,EffLine}. This modification is consistent for all $\gamma$ and is
therefore generic. From this perspective, the modification of
\cite{SphSymmComplex}, which preserves the brackets, requires
$\gamma^2=\epsilon$ and is not generic; it does not provide a way to avoid
non-classical space-time structures without fine-tuning. Our derivations have
shown that the different outcomes of \cite{SphSymmComplex} versus
\cite{JR,HigherSpatial} are not a consequence of working with self-dual
connections (used in \cite{SphSymmComplex}) or real variables (used in
\cite{JR,HigherSpatial}). The crucial difference is that modified constraints
with unmodified brackets, as in \cite{SphSymmComplex}, can be obtained only
for specific $\gamma$, while modifications of constraints as well as brackets
exist for all $\gamma$.

\section{Conclusion}

We have shown that deformations of the classical space-time structure appear
generically in spherically symmetric models of loop quantum gravity. For
self-dual variables or Euclidean gravity with $\gamma=\pm 1$, we have derived
the most general form of the quadratic Hamiltonian constraint free of triad
derivatives, such that a system with unmodified closed brackets is
obtained. This rigidity result, just like the setting of \cite{SphSymmComplex}
which it generalizes, relies on the absence of derivative terms of the triad.
However, from the point of view of an effective field theory, this result is
not generic because it depends on a restriction of derivative terms even
within the  classical structure of second-order derivatives.
Moreover, this rigidity result can be obtained only for specific values of the
Barbero--Immirzi parameter $\gamma$.

The results of \cite{SphSymmComplex} have sometimes been interpreted as saying
that deformations arising in the hypersurface-deformation brackets, obtained
originally using holonomy modifications in real-valued variables, might be
avoided in the self-dual case. Self-dual variables represent a specific choice
for the Immirzi parameter, and therefore do not lead to generic results. These
variables (or the values of $\gamma$ they correspond to) are not distinguished
intrinsically by symmetries bcause constraint brackets, which define the
symmetries of a canonical theory, can be closed for any $\gamma$.

Moreover, we have shown that the possibility of modifications, even
within a self-dual setting, formally depends on the derivative
structure which can be changed by adding multiples of the Gauss
constraint or its spatial derivatives to the Hamiltonian constraint.
This ambiguity can be eliminated by solving the Gauss constraint
explicitly, following \cite{SphSymmHam}, in which case the same
derivative structure is obtained in self-dual type variables and in
real variables, which agrees with the form originally used in an
analysis of modified brackets \cite{JR,HigherSpatial}.  We therefore
conclude that modified brackets and non-classical space-time
structures are generic in any spherically symmetric model with
holonomy modifications, even for self-dual variables. We also pointed
out that currently known modifications may not be generic from the
point of view of canonical effective theory introduced here: After
translating momenta into time derivatives, different derivative orders
appear in the terms of a modified Hamiltonian constraint. This
observation suggests that there is room for further explorations of
possibly new models. A likely candidate for a generic extension is the
inclusion of canonical quantum back-reaction effects
\cite{EffAc,Karpacz,HigherTime}, which in an action formulation
provide higher-curvature terms with generic higher
derivatives. However, quantum back-reaction on its own does not modify
the hypersurface-deformation brackets of constraints \cite{EffConsQBR}
and is therefore unlikely to change our conclusions about modified
space-time structures.

Euclidean and self-dual type variables are special also in an analysis of
cosmological perturbations \cite{CosmoComplex,LQCScalarEucl}, in which case
non-generic modifications of constraint brackets have been observed as
well. Our results present useful indications for operator calculations
\cite{TwoPlusOneDef,TwoPlusOneDef2,AnoFreeWeakDiff,AnoFreeWeak,ConstraintsG,DiffeoOp,OffShell}
which have demonstrated the possibility of off-shell closure of commutators of
constraint operators, mainly in the Euclidean case. So far, these
investigations have not yet given rise to indications that the commutators of
constraint operators may be modified, in contrast to effective derivations as
well as the operator constructions in \cite{ThreeDeform,SphSymmOp}. (However,
it is not always clear how to read off modifications of structure functions in
the operator setting, which should be some function of a spatial metric or
densitized triad and therefore requires a suitable notion of
  states of a semiclassical geometry which does not yet exist in the operator
  formulation.)  Our results show that the Euclidean setting is, in fact,
inconclusive as regards modifications of structure functions because it is a
non-generic case that allows closed brackets with and without
modifications. Current effective and operator treatments are therefore
consistent with one another. For a complete picture of space-time structures
in loop quantum gravity, it will be important to extend off-shell operator
calculations to the full Lorentzian constraint.

\section*{Acknowledgements}

This work was supported in part by NSF grant PHY-1607414 and
PHY-1912168. SB is supported in part by funds from NSERC, from the
Canada Research Chair program and by a McGill Space Institute
fellowship.  MR thanks the Institute for Gravitation and the Cosmos
and Penn State University for the hospitality during the fist stages
of this work. The work of MR has been benefited from partial support
by a mobility grant awarded by Sapienza University of Rome and MIUR.
The contribution of MR is based upon work from COST Action MP1405
QSPACE, supported by COST (European Cooperation in Science and
Technology).

\appendix

\section{Restrictions on coefficients of semi-symmetric Gaussian terms}
\label{appendix-pdesol}
We list the solutions to partial differential equations resulting from the
$\mathcal{H}$-$\mathcal{G}$ and $\mathcal{H}$-$\mathcal{D}$ brackets. These
will give us the so called \textit{semi-symmetric Gaussian} terms. Denoting
$(E^{\varphi})^2=E^{22}+E^{33}$, for $\beta^{ij}$ we have
\[
\left\{
	\begin{aligned}
		&\beta^{11}=\beta^{11}(E^1)\\
		&\beta^{12}=E^3\tilde{C}_{\beta}(E^1)+E^2\bar{C}_{\beta}(E^1)\\
		&\beta^{13}=E^3\bar{C}_{\beta}(E^1)-E^2\tilde{C}_{\beta}(E^1)
	\end{aligned}
	\right.
\]
\[
	\left\{
		\begin{aligned}
			&\beta^{22}=1/2[-8\bar{C}_{\beta^{23}}(E^1)E^{23}
				+(C_{\Sigma}(E^1)+\tilde{C}_{\beta^{23}}(E^1))E^{22}
			+(C_{\Sigma}(E^1)-\tilde{C}_{\beta^{23}}(E^1))E^{33}]\\
			&\beta^{33}=1/2[8\bar{C}_{\beta^{23}}(E^1)E^{23}
				+(C_{\Sigma}(E^1)+\tilde{C}_{\beta^{23}}(E^1))E^{33}
			+(C_{\Sigma}(E^1)-\tilde{C}_{\beta^{23}}(E^1))E^{22}]\\
			&\beta^{23}=\tilde{C}_{\beta^{23}}(E^1)E^{23}+2(E^{22}-E^{33})\bar{C}_{\beta^{23}}(E^1)
		\end{aligned}
			\right.
\]
For $\gamma^i$ we have
\[
\left\{
	\begin{aligned}
		&\gamma^1=\gamma^1(E^1)\\
		&\gamma^2=E^3\tilde{C}_{\gamma}(E^1)+E^2\bar{C}_{\gamma}(E^1)\\
		&\gamma^3=E^3\bar{C}_{\gamma}(E^1)-E^2\tilde{C}_{\gamma}(E^1)
	\end{aligned}
	\right.
\]
For $\alpha^i$ we have
\[
\left\{
	\begin{aligned}
		&\bar{\alpha}^1=C_{\alpha^1}(E^1)E^{\varphi}\\
		&\bar{\alpha}^2=(\tilde{C}_{\bar{\alpha}}(E^1)E^3+\bar{C}_{\bar{\alpha}}(E^1)E^2)E^{\varphi}\\
		&\bar{\alpha}^3=(-\tilde{C}_{\bar{\alpha}}(E^1)E^2+\bar{C}_{\bar{\alpha}}(E^1)E^3)E^{\varphi}
	\end{aligned}
	\right.
\]
\[
\left\{
	\begin{aligned}
&\alpha^1_1=\alpha^1_1(E^1)\\
&\alpha^2_1=E^3\tilde{C}_{\alpha^2_1}(E^1)+E^2\bar{C}_{\alpha^2_1}(E^1)\\
&\alpha^3_1=E^3\bar{C}_{\alpha^2_1}(E^1)-E^2\tilde{C}_{\alpha^2_1}(E^1)\\
&\alpha^1_2=(E^2\tilde{C}_{\alpha_2^1}(E^1)+E^3\bar{C}_{\alpha_2^1}(E^1))\frac{1}{(E^{\varphi})^2}\\
&\alpha^1_3=(-E^2\bar{C}_{\alpha_2^1}(E^1)+E^3\tilde{C}_{\alpha_2^1}(E^1))\frac{1}{(E^{\varphi})^2}
	\end{aligned}
	\right.
\left\{
	\begin{aligned}
		&\alpha^2_2=(-\tilde{C}_{\alpha^2_2}(E^1)E^{23}+\bar{C}_{\alpha^2_2}(E^1)
		E^{33})\frac{1}{(E^{\varphi})^2}\\
		&\alpha^3_3=(\tilde{C}_{\alpha^2_2}(E^1)E^{23}+\bar{C}_{\alpha^2_2}(E^1)
		E^{22})\frac{1}{(E^{\varphi})^2}\\
		&\alpha^2_3=(-\bar{C}_{\alpha^2_2}(E^1)E^{23}+\tilde{C}_{\alpha^2_2}(E^1)
		E^{22})\frac{1}{(E^{\varphi})^2}\\
		&\alpha^3_2=(-\bar{C}_{\alpha^2_2}(E^1)E^{23}-\tilde{C}_{\alpha^2_2}(E^1)
		E^{33})\frac{1}{(E^{\varphi})^2}
	\end{aligned}
	\right.
\]
For $Q$ we have
\[
\left\{
	\begin{aligned}
		&\bar{Q}=(E^{\varphi})^2C_{\bar{Q}}(E^1)\\
		&a_1=E^{\varphi}C_{a_1}(E^1)\\
		&a_2=\frac{E^3}{E^{\varphi}}C_{a_2}(E^1)\\
		&a_3=-\frac{E^2}{E^{\varphi}}C_{a_2}(E^1)
	\end{aligned}
\right.
\left\{
	\begin{aligned}
&c_1=c_1(E^1)\\
&c_2=\frac{E^3}{(E^{\varphi})^2}C_{k}(E^1)\\
&c_3=-\frac{E^2}{(E^{\varphi})^2}C_{k}(E^1)
	\end{aligned}
	\right.
\left\{
	\begin{aligned}
		&b_{11}=b_{11}(E^1)\\
		&b_{12}=(-c_1(E^1)E^2/2+E^3C_b(E^1))\frac{1}{(E^{\varphi})^2}\\
		&b_{13}=(-c_1(E^1)E^3/2-E^2C_b(E^1))\frac{1}{(E^{\varphi})^2}
	\end{aligned}
	\right.
\]

\[
\left\{
	\begin{aligned}
		&b_{22}=(E^{33}C_{b_{22}}(E^1)-3E^{23}C_k(E^1))\frac{1}{(E^{\varphi})^2}\\
		&b_{33}=(E^{22}C_{b_{22}}(E^1)+3E^{23}C_k(E^1))\frac{1}{(E^{\varphi})^2}\\
		&b_{23}=[\frac{3}{2}C_k(E^1)(E^{22}-E^{33})-E^{23}C_{b_{22}}(E^1)]\frac{1}{(E^{\varphi})^2}
	\end{aligned}
	\right.
\]
We also have mixing conditions
\[
\left\{
	\begin{aligned}
		&C_{k}(E^1)=-\gamma^1(E^1)=\tilde{C}_{\alpha^1_2}(E^1)\\
		&C_{a_2}(E^1)=-C_{\alpha^1}(E^1)\\
		&C_b(E^1)=-\frac{1}{2}\alpha^1_1(E^1)\\
		&C_{b_{22}}(E^1)=-\frac{1}{2}\bar{C}_{\alpha^1_2}(E^1)
	\end{aligned}
	\right.
\left\{
	\begin{aligned}
		&\bar{C}_{\alpha^1_2}(E^1)=-2\beta^{11}(E^1)\\
		&-\bar{C}_{\alpha^2_2}(E^1)=2\tilde{C}_{\beta}(E^1)-\bar{C}_{\gamma}(E^1)\\
		&\tilde{C}_{\alpha^2_2}(E^1)=2\bar{C}_{\beta}(E^1)+\tilde{C}_{\gamma}(E^1)
	\end{aligned}
	\right.
\]

\section{Some useful identities}

In calculating the $\{H[N(x)],H[M(x)]\}$ bracket, we can often make use of
antisymmetry and integration by parts to simplify our calculations. Suppose we
only have one canonical pair, then typically we have
\begin{equation}
	H[N(x)]\sim \int {\rm d}x N(x)[\dots+f(E(x),K(x))n(x)+\dots]
	\label{}
\end{equation}
where $n(x)$ is a function of phase-space variables depending on $x$.
Plugging this form of Hamiltonian into the Poission bracket we obtain
non-trivial term
\begin{equation}
	\begin{aligned}
	\{H[N(x)],H[M(x)]\}\ni &\int {\rm d}x {\rm d}y
        \{N(x)M(y)[n(x)\{f(E(x),K(x)),\partial_y^n E(y)\}m(y)]\\ 
        &-(N\leftrightarrow M)\}
	\label{}
\end{aligned}
\end{equation}
Denote $f'(x)\equiv \partial f(E(x),K(x))/\partial K(x)$ and $K_{NM}^{(n)}$
for the above integral term (including the $(N\leftrightarrow M)$), then for
$n=1$ we have
\begin{equation}
	\begin{aligned}
		K_{NM}^{(1)}=-\int {\rm d}x [M'(x)N(x)-N'(x)M(x)]n(x)m(x)f'(x)
	\end{aligned}
	\label{}
\end{equation}

For n=2 we have

\begin{equation}
	\begin{aligned}
		K_{NM}^{(2)}=\int {\rm d}x [M'(x)N(x)-N'(x)M(x)][n(x)f(x)m'(x)-m(x)(n(x)f(x))']
	\end{aligned}
	\label{}
\end{equation}

\end{document}